\documentclass[aps,prx,twocolumn,notitlepage,showpacs,amsmath,amstex,amssymb,citeautoscript,longbibliography,floatfix]{revtex4-2}
\pdfoutput=1
\usepackage[english]{babel}
\usepackage{letltxmacro}
\usepackage{latexsym}
\usepackage{floatrow}
\LetLtxMacro{\ORIGselectlanguage}{\selectlanguage}
\makeatletter
\DeclareRobustCommand{\selectlanguage}[1]{%
  \@ifundefined{alias@\string#1}
    {\ORIGselectlanguage{#1}}
    {\begingroup\edef\x{\endgroup
       \noexpand\ORIGselectlanguage{\@nameuse{alias@#1}}}\x}%
}
\newcommand{\definelanguagealias}[2]{%
  \@namedef{alias@#1}{#2}%
}
\makeatother
\definelanguagealias{en}{english}
\definelanguagealias{English}{english}
\usepackage{graphicx}
\usepackage{amsmath}
\usepackage{braket}
\usepackage{amsfonts}
\usepackage{amssymb}
\usepackage{bm}
\usepackage{color}
\usepackage{soul}
\usepackage{wasysym}
\usepackage{dsfont}
\usepackage{bbold}
\usepackage{bbm}

\usepackage{hyperref}
\hypersetup{
    bookmarks=false,         
    unicode=false,          
    pdftoolbar=false,        
    pdfmenubar=true,        
    pdffitwindow=false,     
    pdfstartview={FitH},    
    pdftitle={Entanglement and precession in two-dimensional dynamical quantum phase transitions},    
    pdfauthor={Stefano De Nicola, Alexios A. Michailidis, Maksym Serbyn},     
    pdfsubject={},   
    pdfcreator={},   
    pdfproducer={}, 
    pdfkeywords={MPS, DQPT, entanglement}, 
    pdfnewwindow=true,      
    colorlinks=true,       
    linkcolor=black,          
    citecolor=blue,        
    filecolor=magenta,      
    urlcolor=blue           
}
\DeclareMathOperator{\Tr}{Tr}

\begin{document}
\title{Entanglement and precession in two-dimensional dynamical quantum phase transitions}
\author{Stefano De Nicola$^{1}$}
\author{Alexios A. Michailidis$^{1,2}$}
\author{Maksym Serbyn$^{1}$}
\affiliation{$^1$IST Austria, Am Campus 1, 3400 Klosterneuburg, Austria}
\affiliation{$^2$Department of Theoretical Physics, University of Geneva, 24 quai Ernest-Ansermet, 1211 Geneva, Switzerland}
\date{\today}
\begin{abstract}
Non-analytic points in the return probability of a quantum state as a function of time, known as dynamical quantum phase transitions (DQPTs), have received great attention in recent years, but the understanding of their mechanism is still incomplete. 
In our recent work~\href{https://doi.org/10.1103/PhysRevLett.126.040602}{[PhysRevLett.126.040602]}, 
we demonstrated that one-dimensional DQPTs can be produced by two distinct mechanisms, namely semiclassical precession and entanglement generation, leading to the definition of precession (pDQPTs) and entanglement (eDQPTs) dynamical quantum phase transitions.
In this manuscript we extend and investigate the notion of p- and eDQPTs in two-dimensional systems by considering semi-infinite ladders of varying width. 
For square lattices, we find that pDQPTs and eDQPTs persist and are characterized by similar phenomenology as in 1D: pDQPTs are associated with a magnetization sign change and a wide entanglement gap, while eDQPTs correspond to suppressed local observables and avoided crossings in the entanglement spectrum. However, DQPTs show higher sensitivity to the ladder width and other details, challenging the extrapolation to the thermodynamic limit especially for eDQPTs. 
Moving to honeycomb lattices, we also demonstrate that lattices with odd number of nearest neighbors give rise to phenomenologies beyond the one-dimensional classification. 
\end{abstract}

\maketitle

\section{Introduction}\label{sec:intro}
In recent years, progress in the development of experimental simulation platforms~\cite{Georgescu2014}, including trapped ions~\cite{blattRoos2012,Schneider_2012} and ultracold atomic gases~\cite{reviewColdAtoms2015,Gross2017}, has opened the door to the study of far-from-equilibrium many-body quantum dynamics~\cite{eisert2015}.
Theoretical and experimental investigations have led to the discovery of a wealth of novel non-equilibrium quantum phenomena, such as the strong~\cite{Nandkishore2015,Abanin2019} or weak~\cite{Turner2018,Serbyn2021} breaking of ergodicity, generalized hydrodynamics describing integrable systems~\cite{doyon2016,bertini2016} and discrete time crystals~\cite{Else2020}. 

A prominent non-equilibrium protocol is the so-called quantum quench~\cite{Calabrese_2005,Calabrese2006,Essler_2016,Mitra2018}, whereby an initial quantum state $\ket{\psi_0}$ is time-evolved with a Hamiltonian $H$ of which $\ket{\psi_0}$ is not an eigenstate. 
Particular interest has been drawn by phenomena occurring on the transient timescales following a quantum quench, which lie within the reach of current experimental and theoretical tools~\cite{Paeckel2019}. Among such phenomena, an analogy between return probabilites of closed quantum systems and equilibrium partition functions has led to the definition of \textit{dynamical quantum phase transitions} (DQPTs)~\cite{heyl2013,heyl2018}. DQPTs are defined as non-analytic points in the time evolution of the return probability (also known as the {fidelity} or {Loschmidt echo} in this context~\cite{heyl2013}).
Since their early discovery in the free-fermion solvable transverse-field Ising chain~\cite{heyl2013}, the existence of DQPTs has been reported in a wide range of models, see e.g.\ Refs~\cite{KarraschSchuricht2013,canovi2014,vajnaDora2015,schmittKehrein2015,Homrighausen2017,zunkovic2018,Gurarie2019,Huang2019,Hagymasi2019} or Ref.~\cite{heyl2018} for a review. Theoretical findings were soon followed by experimental observations on different quantum simulator platforms~\cite{jurcevic2017,Flaschner2018,Tian2018,Guo2019,Wang2019}.

Since DQPTs have first been reported, substantial theoretical effort has been devoted to understanding the conditions underlying their occurrence and their relation to the other physical quantities characterizing a system.
Early observations of DQPTs when quenching across a quantum critical point led to the conjecture of a general relation between DQPTs and ground-state quantum phase transitions; however, while said behavior is frequently observed~\cite{heyl2013,KarraschSchuricht2013,Torlai2014,Karrasch2017,heyl2018}, several counterexamples are known, including both quenches within a phase that nonetheless result in DQPTs and quenches across a quantum critical point that do not~\cite{andraschkoSirker2014,vajnaDora2014,Sharma2015,schmittKehrein2015,jafari2019,heyl2019}.
The relation of DQPTs to local observables has also been widely studied. 
While in free-fermionic models~\cite{heyl2013} or systems with broken symmetries~\cite{heyl2014,weidinger2017,zunkovic2018,feldmeierPollmannKnap2019} DQPTs were shown to be associated with zeros of an order parameter, in other scenarios this relation was found to only hold approximately~\cite{Fogarty_2017,yu2021correlations} or to be absent altogether~\cite{Fogarty_2017}. 
DQPT were also recently studied in relation to suitably defined local string observables, which provide an alternative route to measure their location~\cite{halimeh2020local,Bandyopadhyay2021}.

Similarly, a number of studies sought to identify a connection between DQPTs and the behavior of the entanglement entropy. 
Again, a uniform pattern failed to emerge, with DQPTs in different models being in turn associated with rapid entropy growth~\cite{jurcevic2017}, local maxima in the entanglement entropy~\cite{schmittHeyl2018}, entanglement spectrum crossings~\cite{CanoviPRB2014,Torlai2014,Surace2020} or transitions in a suitably defined entanglement echo~\cite{poyhonen2021entanglement}.
Other works have highlighted a relationship between DQPTs and quasi-particle excitations~\cite{jafari2019,Halimeh2020}. 
Furthermore, within a recently developed stochastic formulation of quantum dynamics~\cite{hoganChalker,ringelGritsev,stochasticApproach}, connections were made between DQPTs and the distribution of the classical stochastic variables encoding the quantum evolution~\cite{stochasticApproach,nonEquilibrium} or the behavior of the saddle point trajectory of an effective action~\cite{sdn2021}.

This complicated picture is compounded by the fact that the current understanding of DQPTs is largely based on the study of one-dimensional systems. The development of a unified framework is even more challenging in higher dimensions, where investigations are hindered by the intrinsic limitations of existing analytical and numerical methods~\cite{Schuch2007,Czarnik2019}.
To this date, studies of DQPTs in higher-dimensional settings have predominantly focused on solvable scenarios, such as the Jordan-Wigner- solvable 2D Kitaev honeycomb~\cite{schmittKehrein2015} and extended toric code~\cite{srivastav2019} models,  the integrable 2D topological Haldane model~\cite{bhattacharya2017}, or quenches in the 2D Ising model for which the rate function can be mapped to the classical Onsager partition function~\cite{Onsager44,heyl2015}.
Mean-field solvable limits have also been considered, including the Falicov-Kimball model~\cite{canovi2014} or the three-dimensional $O(N)$ model for large $N$~\cite{weidinger2017}.
More generally, in the absence of exact solutions a number of numerical methods have been applied. While exact diagonalization can be used to study small systems~\cite{feldmeierPollmannKnap2019}, an important step in the direction of addressing the thermodynamic limit was the development of numerical methods to simulate semi-infinite systems~\cite{James2015,Hashizume2020}.
Studies of two-dimensional systems confirmed the complex picture found in 1D, whereby e.g.\ DQPTs are often but not necessarily associated with ground state phase transitions, but also showed that additional possibilities are present for $D>1$, such as the presence of discontinuities in higher derivatives of the rate function~\cite{heyl2015,schmittKehrein2015}.

Thus, while a large number of studies have revealed that DQPTs are associated with complex, multi-faceted phenomenology, a general understanding of these phenomena has not been attained.
In our recent work~\cite{entanglementView} we demonstrated that a perspective route to understanding one-dimensional DQPTs is given by considering the physical mechanisms leading to their appearance. 
These mechanisms can be revealed by working within the matrix product state (MPS) formalism, which allows one to single out the contributions of semiclassical precession and entanglement generation to the return probability. 
This led to the definition of \textit{precession}- and \textit{entanglement}-driven DQPT (pDQPTs and eDQPTs respectively) to characterize the cases where one of these mechanisms is prevalent. 
The relative importance of the different mechanisms leading to a DQPT is signaled by a number of experimental probes, such as local magnetization and mutual information~\cite{entanglementView}.
When the two mechanisms are simultaneously significant, the resulting DQPT phenomenology is complex and eludes a simple characterization.

In this manuscript we generalize the above picture to lattices of finite width (ladders) that can be extrapolated to two-dimensional spin systems, assessing the stability of p- and eDQPTs and investigating the new possibilities opened up by varying the number of nearest neighbors.
We focus on a semi-infinite cylindrical geometry, previously considered in Refs.~\cite{James2015,Hashizume2020}, which can be treated by MPS methods, allowing a direct generalization of our earlier work.
For square lattices, we show that the local physics following these quenches corresponds to the paradigm of p- and eDQPTs introduced in 1D. Nonetheless, the resulting DQPT phenomenology can present significant differences from the one-dimensional case. 
Furthermore, going beyond square lattices, we demonstrate that connectivity effects can give rise to new possibilities that go beyond the p- vs eDQPT paradigm, while still being understandable in simple terms.

The manuscript is structured as follows. In the next Section we introduce the general framework of DQPTs, outline our MPS-based approach and motivate the notions of p- and eDQPTs. Sections~\ref{sec:precession} and~\ref{sec:entanglement} are respectively devoted to the strong-field and strong-interaction regimes, while in Section~\ref{sec:connectivity} we investigate the new possibilities opened up by changing the ladder connectivity. We conclude in Section~\ref{sec:conclusion} by summarizing our findings and discussing directions for further developments.

\section{MPS formalism for DQPTs}
\subsection{DQPTs in the two-dimensional Ising model}
Consider a time-evolved state $\ket{\psi(t)}=e^{-iHt}\ket{\psi_0}$, where the initial state $\ket{\psi_0}$ is chosen not to be an eigenstate of the Hamiltonian $H$.
In particular, here we shall focus on the two-dimensional quantum Ising model
\begin{multline}\label{eq:ising}
H= \sum_{m n} \left( h_x \sigma_{mn}^x +h_z   \sigma_{mn}^z + J_\parallel \sigma^z_{m n} \sigma^z_{m+1 n} \right.
\\ \left. + J_\perp  \sigma^z_{m n} \sigma^z_{m n+1} \right) ,
\end{multline}
where the subscripts $m\in \{ 1,\dots, L_\perp \}$, $n\in \{1,\dots, L_\parallel \}$ respectively denote the longitudinal and transverse dimensions of a two-dimensional lattice with total number of spins $N=L_\parallel \times L_\perp$. 
We consider periodic boundary conditions along the transverse dimension.

Non-analytic points in the fidelity (return probability) $P(t)=|\langle \psi_0 | \psi(t) \rangle |^2$ following  a quantum quench have been termed DQPTs~\cite{heyl2013}. While $P(t)$ is exponentially suppressed as a function of the number of spins $N$, the fidelity density
\begin{align}
f(t) = -\frac{1}{N} \log | \langle \psi_0 |  \psi(t) \rangle |^2
\end{align}
has a well-defined thermodynamic limit and is therefore the central object of study in this context. 
The appearance of DQPTs in $f(t)$ requires the thermodynamic limit $N \rightarrow \infty$~\cite{heyl2013}. 
Experimentally, this limit can be gradually approached by considering suitably defined local projectors, as discussed in Appendix~\ref{app:local}.
In this manuscript, we consider a semi-infinite cylindrical geometry by coupling $L_\perp$ one-dimensional chains of length $L_\parallel$ and setting $L_\parallel$ to infinity explicitly. The total number of spins $N$ is thus infinite, so that \textit{bona fide} DQPTs can occur even for finite transverse dimension $L_\perp$. 
Below, we shall both investigate the DQPTs observed for finite $L_\perp$ and discuss their stability as $L_\perp$ is increased and the isotropic two-dimensional thermodynamic limit is approached.

\begin{figure}[t]
\begin{center}
\includegraphics[width=0.99\columnwidth]{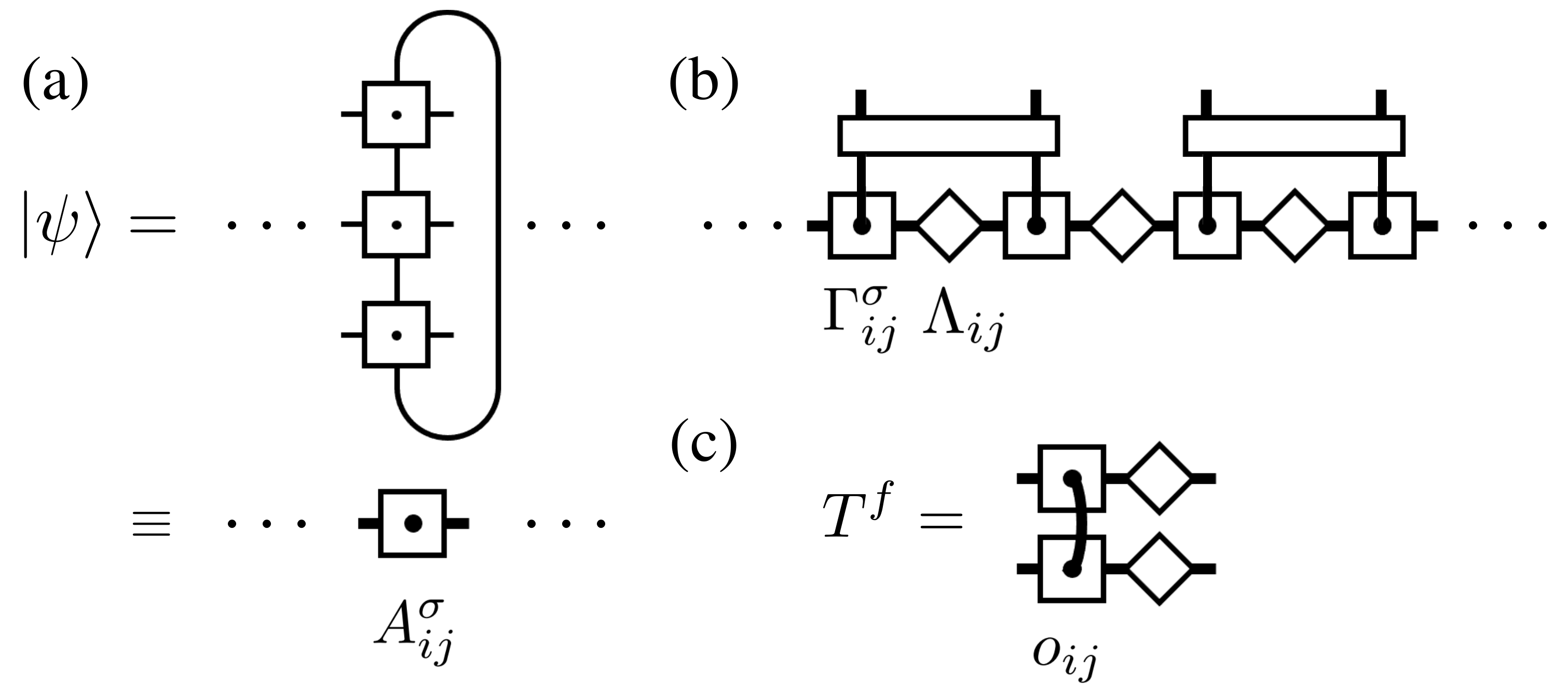}
\end{center}
\caption{\label{fig:method}
(a) MPS representation of a state for a 2D system with a semi-infinite ($\infty \times L_\perp$) cylinder geometry and periodic boundary condition along the finite dimension (here we show $L_\perp=3$). The local state is initially given by a PEPS; in the case of a product state, this is a trivial PEPS with unit bond dimension. By contracting the PEPS along the finite dimension, one obtains an MPS. The thick lines represent physical and virtual indices corresponding to an entire column of the system. 
(b) The MPS can be evolved using the standard iTEBD algorithm retaining the Vidal canonical form, so that the elements of $\Lambda$ encode the entanglement spectrum relative to a transverse bipartition of the system. 
(c) As in 1D, the fidelity is given by the leading eigenvalue of the relevant transfer matrix $T^f$, which is obtained by contracting the time-evolved local state with its conjugate at $t=0$; see Eq.~\eqref{eq:TM}. 
}
\end{figure}

\subsection{Transfer matrix and Vidal canonical form}
The main computational tool used in this work is an infinite MPS (iMPS) representation of the two-dimensional quantum state, similar to that used in~\cite{Hashizume2020}. 
The benefits of this formulation are threefold. 
First, the availability of the canonical gauge (discussed below) allows one to extract the contributions of precession and entanglement generation to the return probability. 
Second, the MPS formulation allows a direct connection to the one-dimensional case, making it easier to identify analogies and differences.
Third, the inherently one-dimensional iMPS encoding (as long as $L_\perp = O(1)$) does not suffer from the complications of simulating higher dimensional tensor networks.

The iMPS representation used in this manuscript is shown in Fig.~\ref{fig:method}.
This construction is inspired by a projected-entangled pair state (PEPS) with initial bond dimensions $\chi^{\parallel}_0$, $\chi^{\perp}_0$, respectively along and across the chains, and physical dimension $d$. 
The simplest case of this is given by an initial product state, $\chi^\parallel_0 = \chi^\perp_0 =1$, which we focus on in our numerical analysis.
The MPS representation of the state at $t=0$ is obtained by contracting along the finite perpendicular dimension. 
This yields an MPS written in terms of a $\chi \times d^{L_\perp} \times \chi$ local tensor $A^\sigma_{ij}$ (where we denote {$(\chi^\parallel_0)^{L_\perp}=\chi$}), represented by a square in the bottom part of Fig.~\ref{fig:method}(a). 
The local tensor elements $A^\sigma_{ij}$ are labeled by  two virtual indices, $i,j \in \{ 1,\ldots, \chi \}$, and a physical index, $\sigma$. 
The index $\sigma$ runs over all possible local spin configurations, e.g. for spin-$1/2$ the one-dimensional case $L_\perp=1$ corresponds to $\sigma \in \{ \uparrow, \downarrow \}$, while for general $L_\perp$ the index $\sigma$ runs over all possible tensor products featuring $L_\perp$ copies of either $\{ \uparrow, \downarrow \}$. 

This MPS can then be time-evolved using infinite time-evolving block decimation (iTEBD)~\cite{Vidal2006}. 
In practice, this amounts to the repeated application of pair-entangling gates, as shown in Fig.~\ref{fig:method}(b). 
Further details on the practical implementation of iTEBD used in this manuscript are given in Appendix~\ref{app:computational}.
The state thus evolved retains the \textit{canonical form}~\cite{Vidal2006,Orus2008} 
\begin{equation}\label{Eq:canonical}
A^\sigma_{ij}(t) =  \Lambda_{ii}(t) \Gamma^\sigma_{ij}(t),
\end{equation} 
where the diagonal matrix $\Lambda_{ij}(t) = \delta_{ij} \sqrt{\lambda}_i$ features the singular values $\lambda_i$ of the Schmidt decomposition with respect to a horizontal bond. 
The ordered singular values, $\lambda_i \geq\lambda_{i-1}$, constitute the entanglement spectrum for a transverse bipartition of the system, i.e perpendicular to the chains, and therefore the normalization of the state corresponds to $\sum_i \lambda^2_i = 1$.   The corresponding bipartite entanglement entropy is then given by $S=-\sum_i \lambda_i \log \lambda_i$.
The remaining tensor $\Gamma_{ij}^\sigma$ carries a physical index, so that its elements $\Gamma_{ij} = (\Gamma_{ij}^{\uparrow \uparrow \dots \uparrow},\Gamma_{ij}^{\downarrow \uparrow \dots \uparrow},\cdots ,\Gamma_{ij}^{\downarrow\downarrow \dots \downarrow})$ can be viewed as (not necessarily normalized) quantum states. 
The tensors $\Lambda$ and $\Gamma$ satisfy the canonical conditions
$\sum_{ij\sigma} \Lambda_{ij}^2 \Gamma^{\sigma}_{jk} \Gamma^{\sigma *}_{il}=  \sum_{ij\sigma} \Lambda_{ij}^2 \Gamma^\sigma_{kj} \Gamma^{\sigma *}_{li}  = \delta_{kl}$~\cite{
Orus2008}. 

The MPS representation of the state makes it straightforward to compute the fidelity for a translationally invariant semi-infinite system. 
Namely, the fidelity is obtained from the leading eigenvalue of the \textit{fidelity transfer matrix} $T^f$, computed by contracting the local tensors corresponding to the initial and time-evolved states~\cite{andraschkoSirker2014,piroli2018,entanglementView}. 
Given the spectrum $\{ e_i \}$ of $T^f$, one has
\begin{align}
f=-\frac{ 2}{L_\perp} \log  \max_i(| e_i|  ).
\end{align}
At $t=0$, normalization imposes $e_1=1$, $e_{i} =0 $ $\forall \, i\neq 1$. Following a quantum quench, the spectrum $\{ e_i \}$ smoothly evolves in the complex plane, with DQPTs occurring whenever the subleading eigenvalue $e_2$ overtakes $e_1$ in magnitude.
For initial product states $\ket{\psi_0} = \otimes_m\ket{v}_m$ one has $\Gamma^{\sigma}_{ij}(0)= v^\sigma$, so that  $T^f_{ij}(t) =\sum_\sigma  [v^\sigma]^* A^{\sigma}_{ij}(t)  = \braket{v|A_{ij}(t)}$.
This is pictorially shown in Fig.~\ref{fig:method}(c).
By construction, the transfer matrix $T^f$ is made up of two contributions, 
\begin{align}\label{eq:TM}
T^f_{ij}= o_{ij} \Lambda_{ii}.
\end{align}
Here $o_{ij} =  \braket{v|\Gamma_{ij}}$ is a matrix of  overlaps, which we shall discuss further below, and $\Lambda$ encodes the contribution of the entanglement spectrum.

\subsection{Precession and entanglement DQPTs}

The significance of the contributions $o_{ij}$ and $\Lambda_{ii}$ to the fidelity can be understood by considering two limiting cases.
First, let us consider the  dynamics induced on a generic initial product state $\ket{\psi_0}=\otimes_m \ket{v}_m$ by the Hamiltonian~(\ref{eq:ising}) in the non-interacting limit $J_\perp=J_{\parallel}=0$. 
Since no entangling terms are present, the system remains in a product state at all times. 
The bond dimension is then unity, so that $\Lambda=1$ and $\Gamma = \ket{\Gamma_{11}}= \ket{v(t)}$ is the time-evolved local state, which performs precession under the action of the time evolution operator.
The leading eigenvalue of the transfer matrix $T^f$ is then immediately given by $|e_1|=|o_{11}|= |\braket{v(t)|v(0)}|$, so that (trival) DQPTs in the system can occur whenever precession leads $\ket{v(t)}$ to be orthogonal to $\ket{v(0)}$.
In the presence of finite but comparatively small interactions, the entanglement spectrum becomes non-trivial but a large entanglement gap persists, $\lambda_1 \gg \lambda_2 $. Thus, $\Gamma_{11}$ is still the dominant contribution to the state. DQPTs occur near minima of $|o_{11}|$, where $\Gamma_{11}$ has rotated maximally away from the initial state so that the magnetization would take an opposite expectation value to that in the initial state. This phenomenology, which amounts to semiclassical precession with corrections given by $\lambda_i$ with $i>1$, corresponds to pDQPTs~\cite{entanglementView}.

In contrast, consider the exactly solvable case of a one-dimensional chain ($J_\perp=0$) initialized in the $\ket{\psi_0}= \otimes_m \ket{\rightarrow}_m$ state, which is the ground state of \eqref{eq:ising} for $h_x \rightarrow -\infty$, and evolved with the Hamiltonian~\eqref{eq:ising} with $J_\parallel =  J \neq 0$, $h_x=h_z=0$. In this case, it can be shown~\cite{entanglementView} that the transfer matrix $T^f= o \Lambda$ is exactly given by the product of a $\Lambda$ matrix that has $\sqrt{\lambda}=\{ |\cos(J t)|, |\sin (J t)| \}$ on the diagonal and a simple matrix of overlaps that does not depend on time, 
\begin{align}\label{eq:overlapMatrix_classical}
o =\left(\begin{matrix} 1 &  0 \\
0 &  -i  \end{matrix} \right), 
\end{align}
so that the leading eigenvalue of $T^f$ reads $|e_1(t)|=\max  \{ |\cos(J t)| , |\sin (J t)| \} $. 
Thus, as the overlap matrix is constant, DQPTs in this regime are entirely determined by crossings in the entanglement spectrum. 
If a finite but small external field is present, the overlaps $o_{ij}$ show slow time evolution, and DQPTs are still predominantly driven by (avoided) crossings in the entanglement spectrum. Due to the large entanglement, local expectation values near such DQPTs are typically suppressed. This phenomenology corresponds to eDQPTs.

In more general scenarios, both the mechanisms described above will be simultaneously present; however, in many relevant settings only one of them is found to be predominant. 
When instead both mechanisms contribute to a comparable extent, their competition gives rise to a complex intermediate regime, where DQPTs can be very sensitive to the quench details and can escape a simple characterization in terms of the overlaps, entanglement spectrum or local observables~\cite{entanglementView}.

In the following sections, we shall investigate whether the described p- and eDQPTs phenomenology persists in the case of ladders that upon increasing their width approach two-dimensional systems. 
To do so, we shall consider quantum quenches in the strong-field and strong-interaction regimes, which, based on the behavior of 1D systems, would be expected to give rise to p- and eDQPTs respectively.

\section{Strong-field regime \label{sec:precession} }

\begin{figure}[t]
\begin{center}
\includegraphics[width=0.8\columnwidth]{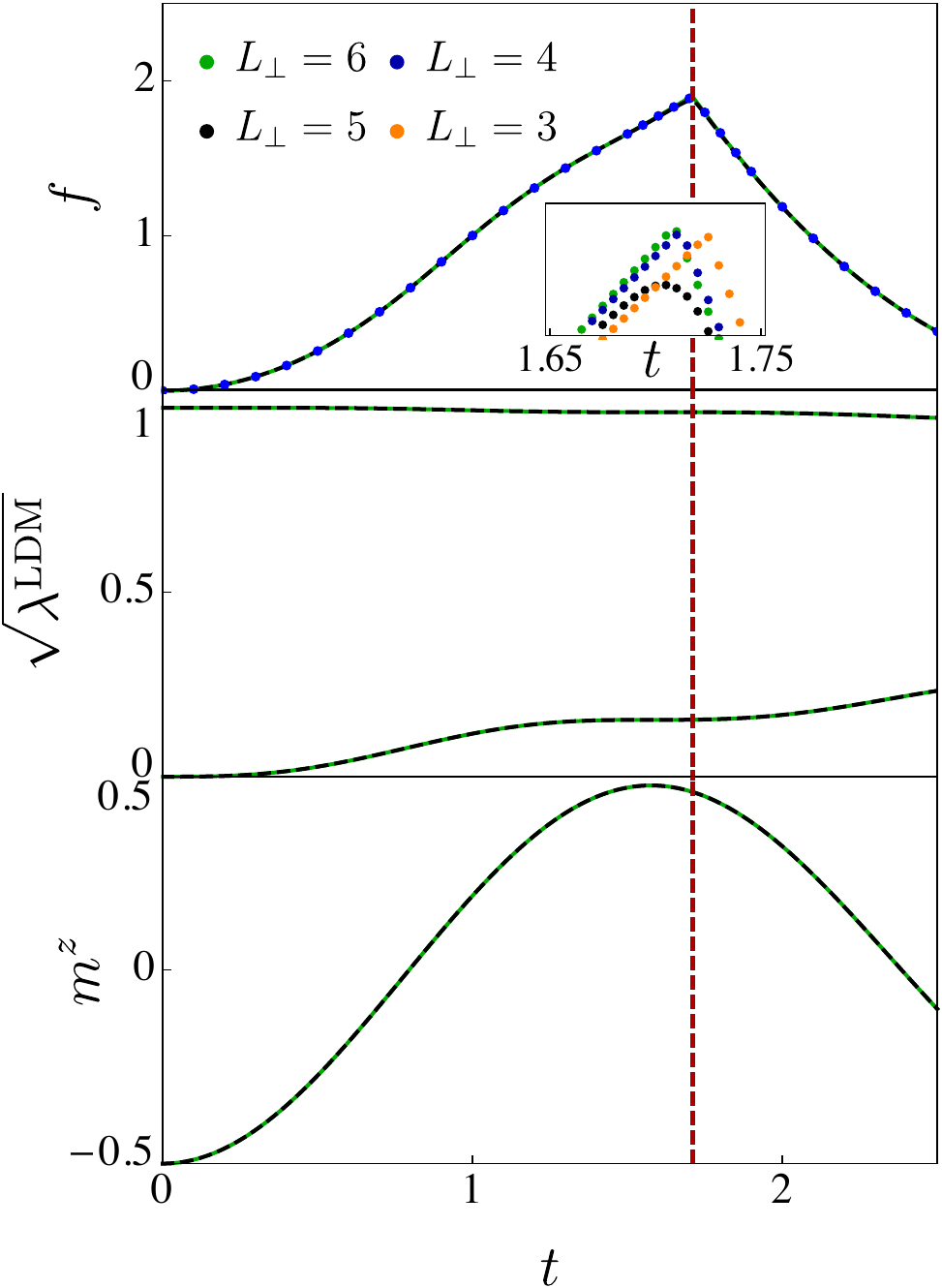}
\end{center}
\caption{\label{fig:pDQPT}
For quantum quenches where strong precession-inducing terms are present, we expect the appearance of pDQPTs. 
Here we select the initial state $\ket{\psi_0}=\otimes_m \ket{\downarrow}_m$ and evolve using~\eqref{eq:ising} with $h_x=1$, $h_z=0$, $J_\parallel = J_\perp=0.1$.
For most of the transverse sizes $L_\perp$ we consider, a DQPT is indeed observed, and the fidelity density $f$ appears to rapidly converge with $L_\perp$; this is shown by a comparison of $L_\perp = 4,5,6$ in the top panel. 
However, the DQPT is not present for $L_\perp =5$, as illustrated in the inset which additionally shows $L_\perp=3$. 
This discrepancy is not observed in the behavior of the local entanglement spectrum $\lambda^{\text{LDM}}$ and local magnetization $m^z$ at the DQPT (dashed vertical line), which appear to be converged with respect to $L_\perp$ as demonstrated by comparing $L_\perp=5$ (black dashed lines) and $L_\perp=6$ (green solid lines). 
Their time-evolution follows the same pattern observed for 1D pDQPTs, namely a wide entanglement gap and an inversion of the magnetization. 
}
\end{figure} 

\subsection{Fidelity and local observables}
We begin by considering the strong-field regime, where precession can be expected to dominate the early-time dynamics; this scenario is associated to pDQPTs in one-dimensional systems~\cite{entanglementView}. 
We consider a quantum quench from the state $\ket{\psi_0}=\otimes_m \ket{\downarrow}_m$, which is the ground state of~\eqref{eq:ising} for $J_\parallel, J_\perp \rightarrow -\infty$, $h_z>0$. 
The time-evolution is determined by the Hamiltonian~\eqref{eq:ising} with $J_\parallel=J_\perp=0.1$, $h_x=1$, $h_z=0$.
The resulting dynamics, shown in Fig.~\ref{fig:pDQPT}(top), shows that DQPTs occur for $L_\perp =3,4,6$.

In order to further probe the nature of the observed DQPTs, we use the magnetization in the direction of the initial state, here $m^z(t)=\bra{ \psi(t) } \sum_i S_i^z \ket{\psi(t)}/N$ with $S_i^z = \sigma^z_i/2$, and the one-site entanglement spectrum of a single spin relative to the rest of the system, $\{ \lambda^{\mathrm{LDM}}_i \} $. 
The spectrum $\{ \lambda^{\mathrm{LDM}}_i \}$ is obtained from the one-site reduced density matrix (or local density matrix, LDM), so we refer to it as ``local entanglement spectrum". For the purpose of probing the underlying physics, $\{ \lambda^{\mathrm{LDM}}_i \}$ replaces the entanglement spectrum $\{ \lambda_i \}$ relative to a transverse bipartition of the system (i.e. across the chains), which was considered in 1D~\cite{entanglementView}; this is because the latter does not contain information about entanglement along the transverse dimension.  However, we shall see below that the transverse bipartite entanglement spectrum $\{ \lambda_i \}$ still plays a role in determining the stability of DQPTs.

The middle panel of Fig.~\ref{fig:pDQPT} shows the local entanglement spectrum, which reveals clear pDQPT character manifested by a large gap $\lambda_1 \gg \lambda_2$. The bottom panel shows the magnetization changing its sign at the time of DQPT, $m^z(t) \approx - m^z(0)$. 
Furthermore, the observed behavior of the fidelity, magnetization and local entanglement are qualitatively reproduced by an immediate two-dimensional generalization of the analytical pDQPT ansatz introduced in Ref.~\cite{entanglementView}, which encodes the relevant local physics; see Appendix~\ref{app:ansatz}. 
This further supports the conjecture that the physics of two-dimensional pDQPTs closely mirrors the one-dimensional case.

We however note that for $L_\perp=5$ we do not observe a DQPT but, rather, a smooth peak.
The absence of a DQPT for  $L_\perp=5$, however, does not manifest itself in the dynamics of local observables, which appear to be converged with respect to the system size. 
In fact, the local entanglement and magnetization are nearly indistinguishable for $L_\perp=5,6$. 
We find that the DQPT for $L_\perp=5$ can be restored by the inclusion a small perturbation (e.g. a small rotation of the external field) which does not significantly alter the local physics, as detailed in Appendix~\ref{app:pDQPT}. This suggests that pDQPTs in $D>1$ could be less stable than for $D=1$, as further discussed below.

\subsection{Overlaps and transverse entanglement spectrum}
\begin{figure}[b]
\begin{center}
\includegraphics[width=0.8\columnwidth]{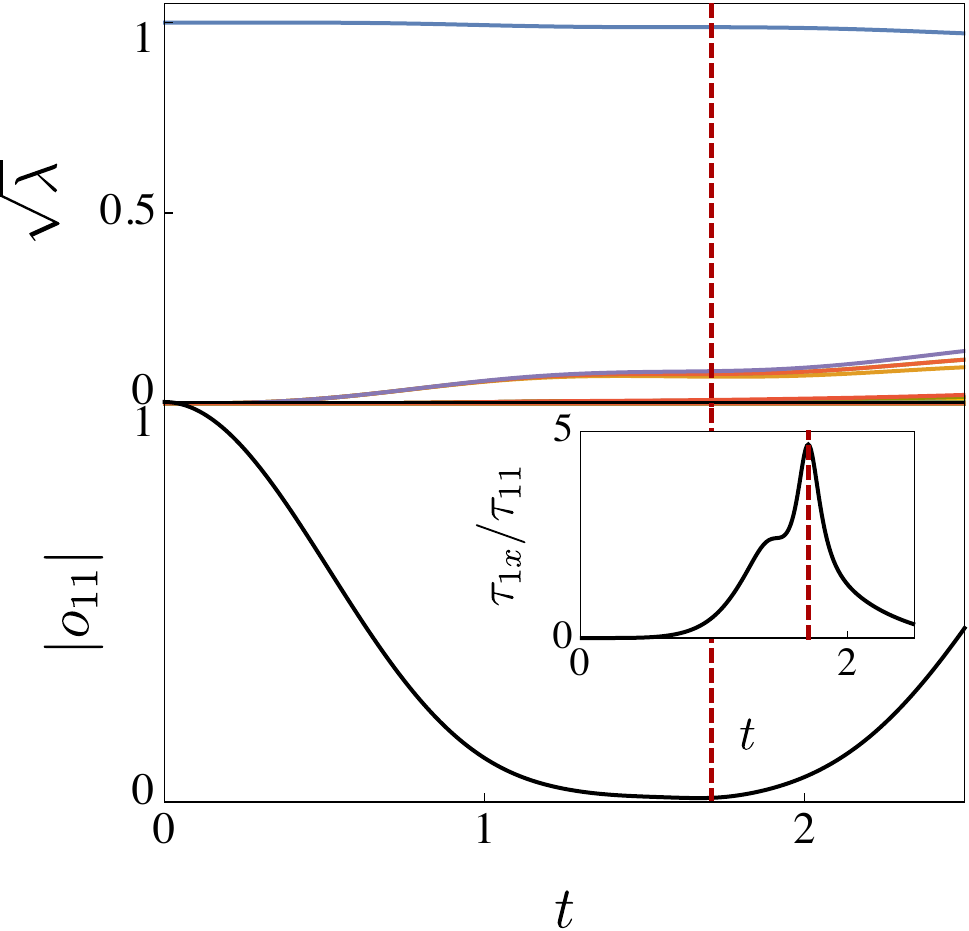}
\end{center}
\caption{\label{fig:pDQPToverlaps} 
{Transverse bipartite} entanglement spectrum $\{ \lambda_i \}$ and overlap $|o_{11}|$ corresponding to the dominant product state for the quench of Fig.~\ref{fig:pDQPT}; here we focus on $L_\perp=4$.
At the DQPT, we observe a large gap between $\lambda_1$ and a small set of singular values $\{ \lambda_2,\ldots, \lambda_5\}$, which are in turn significantly larger than the bulk of the spectrum. 
The DQPT occurs in the vicinity of the minimum of $|o_{11}|$. 
As shown in the inset, this is also near a maximum of the overall relative amplitude $\tau_{1x} /\tau_{11}$ corresponding to the creation of excitations over the dominant product state.
}
\end{figure}

In order to further investigate the nature of the DQPTs we observed, in Fig.~\ref{fig:pDQPToverlaps} we consider the evolution of the dominant overlap $o_{11}$ and the spectrum of transverse bipartite entanglement $\lambda_i$. 
In the presence of a wide entanglement gap, the full quantum state is well-approximated by a product state given by the local tensor $\Gamma_{11}$ at all sites. 
Within a semiclassical picture, the presence of non-zero $\lambda_i<\lambda_1$ can be accounted for in terms of a superposition of the dominant product state with states featuring some ``excitations", i.e. states that differ from the uniform $\Gamma_{11}$ product state by the replacement of some local tensors by various $\Gamma_{ij}$. Each $\Gamma_{11}$ is multiplied by $\lambda_1$, whereas the amplitude for including each $\Gamma_{ij}$ is proportional to $\sqrt{\lambda_i \lambda_j}$; see the supplemental material of~\cite{entanglementView} for further details.
Excitations will then give a correction to the dominant overlap $o_{11}$ obtained from the uniform $\Gamma_{11}$ product state. The excitation of lowest order in $\lambda_j<\lambda_1$ features the tensors $\Gamma_{1j}$, $\Gamma_{j1}$. The corresponding contribution to the overlap is given by
\begin{align}
\tau_{1j} = |o_{1j} o_{j1}| \sqrt{ \lambda_1 \lambda_j}
\end{align}
where $\tau_{11}$ corresponds to no excitations being created. The total contribution from excitations can then be estimated as $\tau_{1x}\equiv \sum_{j\neq 1} \tau_{1j}$.

To illustrate the behavior of overlaps and entanglement, we use the same quench parameters as in Fig.~\ref{fig:pDQPT}. For ease of visualization, we show $L_\perp=4$ since $o_{11}$ and the $\lambda_i$ become smaller for increasing $L_\perp$; however, the same qualitative results hold for the other system sizes we considered.
As in the one-dimensional case, we find that the DQPT occurs in the presence of a wide entanglement gap and near the minimum of $o_{11}$, two signatures of pDQPTs.
Due to the wide gap in $\{ \lambda \}$, the time-evolved state remains close to a product state, so that the overlaps $o_{ij}$ mostly control the onset of DQPTs. 
The inset of Fig.~\ref{fig:pDQPToverlaps} further shows that the DQPT is found to occur in the vicinity of a maximum of the relative excitation amplitude, $\tau_{1x}/\tau_{11}$, further suggesting that the one-dimensional picture featuring excitations over semiclassical precession still holds. 

However, we find that several entanglement eigenvalues $\{\lambda_i\}$ cluster in the vicinity of $\lambda_2$; this is in contrast with the one-dimensional case, where we found $\lambda_1 \gg \lambda_2 \gg \lambda_3$ for pDQPTs.
This observation can be understood by considering the case of $L_\perp$ decoupled parallel chains, $J_\perp=0$. In this scenario, the full system is described by a direct product of one-dimensional systems. Thus, the bipartite entanglement spectrum can be obtained from all possible products featuring $L_\perp$ elements of the one-dimensional spectrum $\{ \lambda^{1D} \}$,
\begin{align}
\label{Eq:ld}
\{ \lambda_i \} = \{ (\lambda_1^{1D})^{a_1} (\lambda_2^{1D})^{a_2} \cdots, \cdots \} 
\end{align}
with $\sum_i a_i = L_\perp$. In this limit, the leading eigenvalue is given by $(\lambda_1^{1D})^{L_\perp}$ and is non-degenerate; there are then $L_\perp$ degenerate subleading eigenvalues, given by $\lambda_2= (\lambda_1^{1D})^{L_\perp-1} \lambda_2^{1D}$. 
In the presence of a non-zero $J_\perp$, this trivial degeneracy is lifted and the entanglement spectrum can no longer be obtained from the one-dimensional one. 
However, the formerly degenerate $L_\perp$ eigenvalues still take similar values, clustering in the vicinity of $\lambda_2$.
Furthermore, some degree of degeneracy is retained, due to additional symmetries for ladders; for instance, for the $L_\perp =4$ system of Fig.~\ref{fig:pDQPToverlaps} we find that the eigenvalues $\lambda_3=\lambda_4$ form a degenerate pair.
This ``concentration'' of singular values in the vicinity of $\lambda_2$ suggests that different comparable contributions could compete, thus complicating the picture with respect to the 1D case; this could explain the sensitivity of DQPTs to the system's details demonstrated by the $L_\perp=5$ case.

In summary, our analysis shows that the one-dimensional picture of pDQPTs in terms of excitations over a leading semiclassical precession also holds in two dimensions.
However, we find evidence that higher-dimensional pDQPTs could be more sensitive to the system's details compared to $D=1$, making it potentially harder to accurately predict their occurrence and to extrapolate to the $L_\perp \rightarrow \infty$ limit.

\section{Strong-interaction regime}
\label{sec:entanglement}

\begin{figure}[tb]
\begin{center}
\includegraphics[width=0.8\columnwidth]{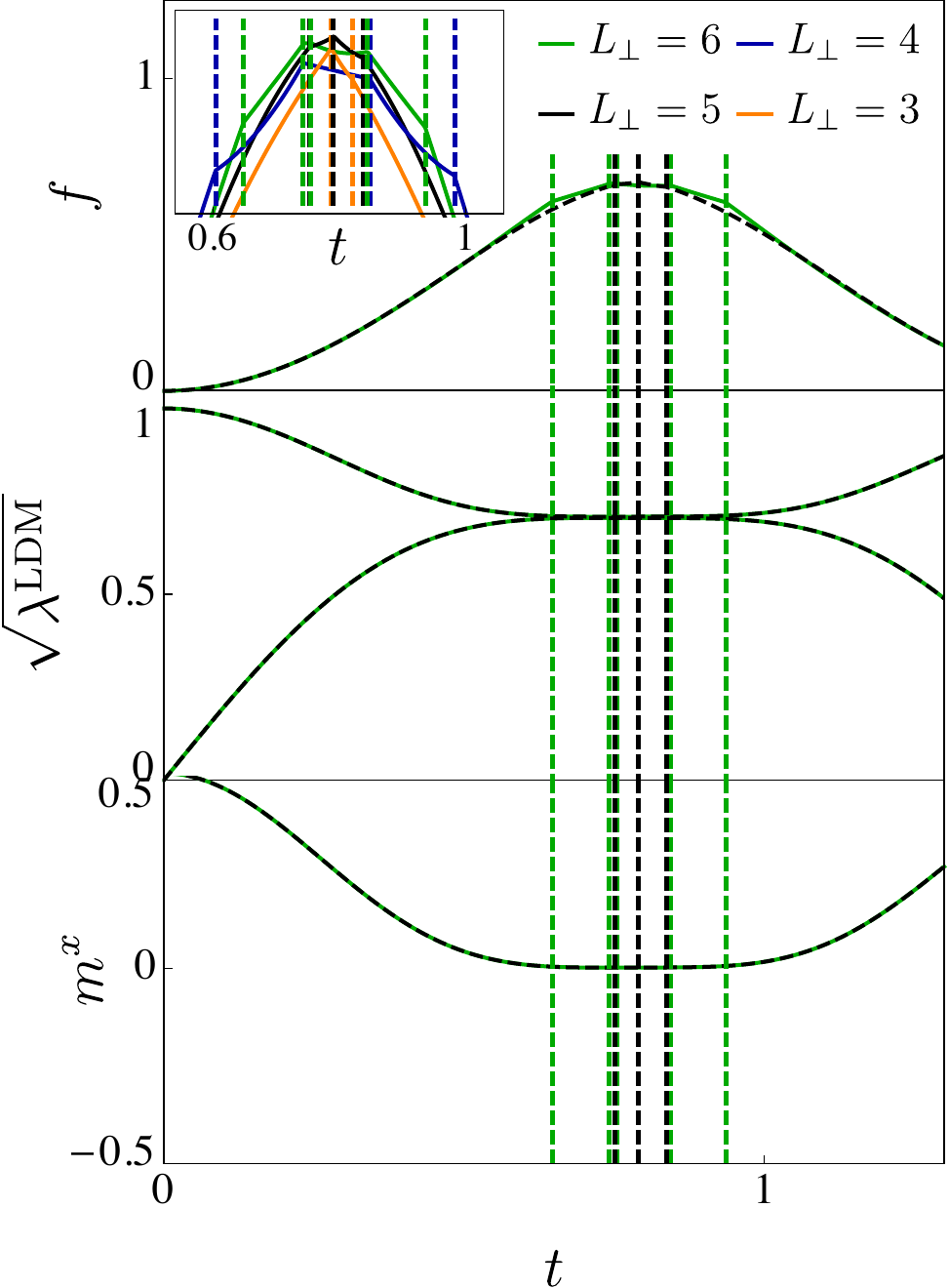}
\end{center}
\caption{\label{fig:eDQPT}
In the presence of strong entanglement-generating terms, we expect to observe eDQPTs. 
Here we consider the initial state $\ket{\psi_0} = \otimes_m\ket{\rightarrow}_m$ evolved using the Hamiltonian~\eqref{eq:ising} with $h_x=0.1$, $h_z=0$, $J_\parallel=J_\perp=1$.
The behavior of the local entanglement and magnetization confirm our expectations based on 1D, with DQPTs occurring near high-entangled regions when local observables are suppressed. 
However, the fidelity $f$ appears to converge slowly with respect to $L_\perp$: in fact, not only the position but also the number of DQPTs (dashed vertical lines) changes upon increasing $L_\perp$, as further illustrated in the inset.
This is in spite of the local physics showing rapid convergence with respect to $L_\perp$, as demonstrated by the two lower panels. 
}
\end{figure}

\subsection{Fidelity and local observables}
Let us now turn to the strong-interaction regime. 
Here we consider a quench from $\ket{\psi_0}=\otimes_m \ket{\rightarrow}_m$, which is the ground state of the Hamiltonian~\eqref{eq:ising} for $h_x \rightarrow -\infty$. We time-evolve using~\eqref{eq:ising} with a weak external field $h_x=0.1$, $h_z=0$ and strong isotropic interactions $J_\parallel= J_\perp=1$. 
In 1D, similar quench parameters were observed to give rise to eDQPTs. However, crucially, one-dimensional interactions only couple the spins along rows, corresponding to $J_\perp=0$ in the present language.
As we shall see, this gives rise to qualitative differences for the case of ladders.

Figure~\ref{fig:eDQPT} shows the suppressed local magnetization $m^x\approx 0$ and the small entanglement gap that are indicators of eDQPT phenomenology. However, the behavior of DQPTs is found to be unstable. Not only the position, but also the number of DQPTs is observed to change as the system size is increased, as is emphasized in the inset of Fig.~\ref{fig:eDQPT}. 
Specifically, although these DQPTs tend to occur at a similar time, we find $2$ DQPTs for $L_\perp=3$, $4$ DQPTs for $L_\perp=4$, $3$ DQPTs for $L_\perp=5$ and $5$ DQPTs for $L_\perp=6$, showing an overall tendency for more DQPTs to arise as $L_\perp$ increases.
Irregular behavior of the number of DQPTs as a function of $L_\perp$ is also reproduced by the 2D generalization of the analytical eDQPT Ansatz introduced in Ref.~\cite{entanglementView}, discussed in Appendix~\ref{app:ansatz}. However, the Ansatz does not match the number and location of DQPTs given by iTEBD, in spite of accurately capturing the local physics for this quench. This further points to the fact that two-dimensional DQPTs are the result of a complex interplay.

In addition to the irregular behavior of the number of DQPTs with $L_\perp$, we find that eDQPTs on finite width lattices can also occur by the leading and subleading eigenvalues of $T^f$ becoming degenerate, rather than crossing, a possibility that we had not encountered in $D=1$. This is illustrated in the top panel of Fig.~\ref{fig:eDQPToverlaps}, where we show $e_{1,2}$ for the quench of Fig.~\ref{fig:eDQPT} and $L_\perp=4$. We checked that $|e_1|-|e_2|$ gets smaller as the bond dimension is increased or the time step is decreased, which points to a true degeneracy between the lowest two eigenvalues of the transfer matrix.

\subsection{Overlaps and transverse entanglement spectrum}

\begin{figure}
\begin{center}
\includegraphics[width=0.8\columnwidth]{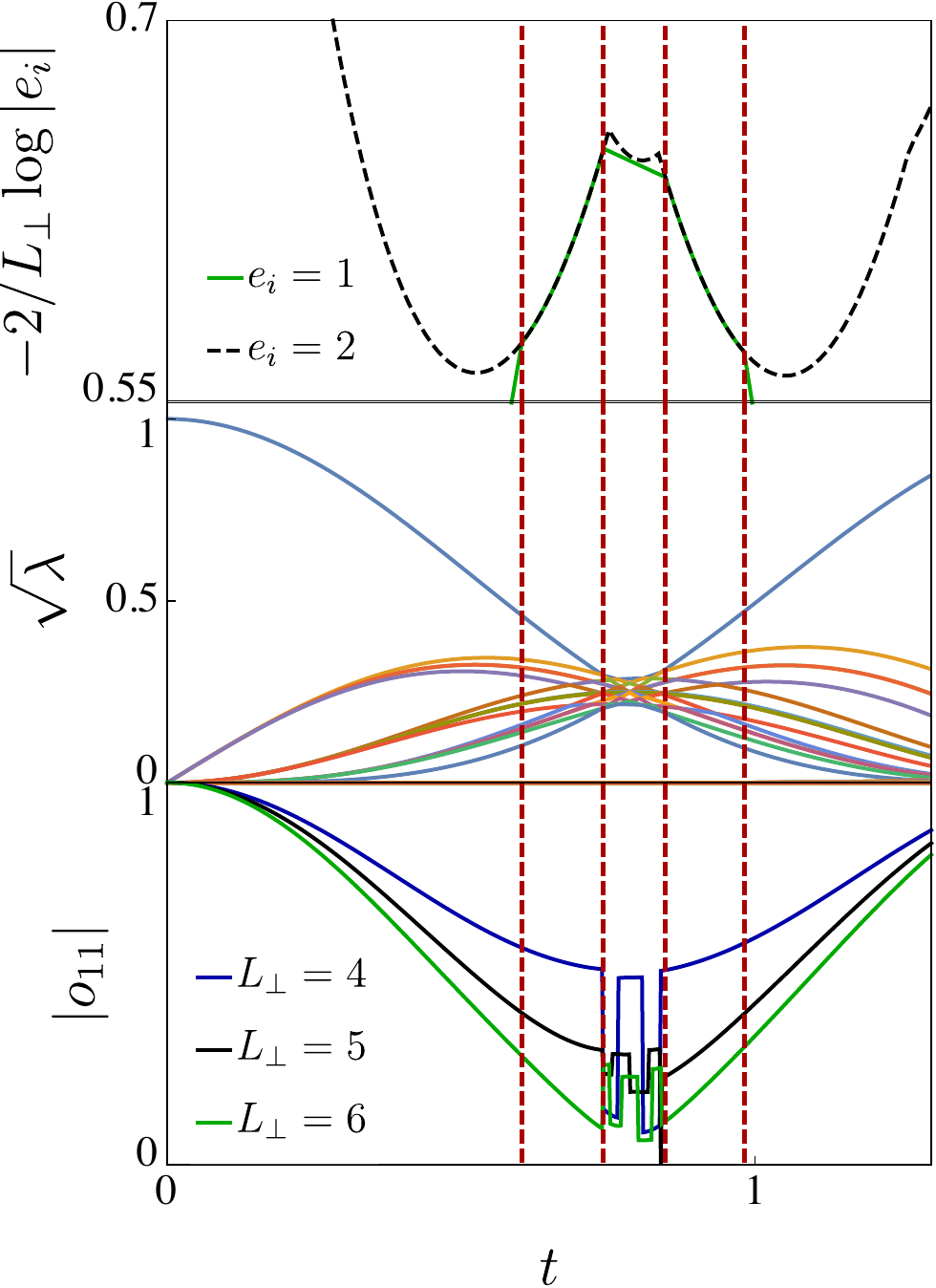}
\end{center}
\caption{\label{fig:eDQPToverlaps} 
Top: In the strong-interaction scenario, corresponding to the quench of Fig.~\ref{fig:eDQPT},
we find that DQPTs can occur by the leading and subleading eigenvalues of $T^f$, $e_1$ and $e_2$, becoming degenerate rather than simply crossing; this is shown in the top panel for $L_\perp=4$. 
In this regime, the entanglement spectrum shows a complicated pattern with several crossings and avoided crossings in the vicinity of the DQPTs, as shown in the middle panel again for $L_\perp=4$. 
Additionally, DQPTs happen near to a minimum of the initially dominant overlap $o_{11}$, which appears to perform precession. This effect becomes more pronounced as $L_\perp$ is increased, as demonstrated in the bottom panel.
These observations entail that both precession and entanglement constitute a significant driving factor for DQPTs, which are then always in a complicated hybrid regime.
The sudden jumps observed in $|o_{11}|$ correspond to crossings between $\lambda_{1}$ and $\lambda_{2}$, where the dominant product state (i.e. associated with $\lambda_1$) changes abruptly. Such crossings occur in the region where there exist several $\lambda_i \approx \lambda_1$ as explained in the main text.
}
\end{figure}

The seeming instability of eDQPTs in $D=2$ can be understood by once again considering the overlaps and transverse bipartite entanglement contributions to the fidelity shown in Fig.~\ref{fig:eDQPToverlaps}.
 While the precise location and number of DQPTs varies depending on system size, they occur in the same time region and the behavior of $\lambda$ and $o$ is qualitatively similar.
In Fig.~\ref{fig:eDQPToverlaps}, we observe that the behavior of entanglement is indeed reminiscent of eDQPTs, with a closing of the transverse bipartite entanglement gap. However, this behavior now involves several $\lambda_i$, which perform multiple crossings and avoided crossings within a small region comprising the DQPTs. 
This high level of near-degeneracy in $\{ \lambda_i \}$ can again be understood as arising from the contribution of the different identical rows, as discussed for pDQPTs, see Eq.~(\ref{Eq:ld}).

In addition, however, the behavior of the overlaps $o_{ij}$ is very different from the case of one-dimensional eDQPTs. 
In one dimension, the $o_{ij}$ show slow evolution, apart from avoided crossings near eDQPTs. 
In contrast, here we observe that the dominating overlap $o_{11}$ performs precession and by the DQPT time $|o_{11}|$ is near a minimum, which is gradually further suppressed as $L_\perp$ is increased.
The cause of this lies in the different meaning of the $o_{ij}$ for $D=2$. By constructions, these are the overlaps between the initial and time-evolved local state for \textit{a whole column} of the system. This means that the overlap matrix $o_{ij}$ actually encodes entanglement along the finite transverse dimension, whereas entanglement in the longitudinal dimension is encoded in the $\lambda_i$.
Consider the case of uncoupled columns of length $L_\perp$, i.e. $J_\parallel=0$. In this case, one has $\Lambda=1$ and the entire dynamics is encoded in the $\Gamma$ matrix. From studies of the one-dimensional Ising chain, we know that DQPTs will occur whereby the time-evolved local state is ``maximally orthogonal" to the initial state; these must therefore correspond to minima of $o_{11}$.
In the presence of comparable longitudinal and transverse interactions, $J_\perp \approx J_\parallel $, we have both a strong contribution of the transverse entanglement spectrum, capturing interactions along the chains, and the overlaps, which capture interactions across the chains.

Thus, DQPTs on ladders and likely in two-dimensional lattices are produced by the outcome of competing contributions. Both overlaps and entanglement are relevant, so that the system is always in the complex intermediate regime identified in Ref.~\cite{entanglementView}, leading to unstable behavior.
This, in addition to the presence of several $\lambda_i \approx \lambda_1$ at the DQPT, further leads to the breakdown of the simple one-dimensional picture describing eDQPTs in terms of a small number of contributions.
Thus, while we still find that it is possible to observe eDQPTs on ladders, characterized by large entanglement and suppressed local observables, it is difficult to produce a simple characterization of their dynamics capable of exactly predicting their occurrence, due to fact that they originate from a number of competing contributions.
These observations point to a non-universal picture for DQPTs in the strong-interaction regime, and makes it challenging in practice to extrapolate to the two-dimensional limit, $L_\perp \rightarrow \infty$.

For both p- and eDQPTs we observed a significant dependence on the system's details, such as the transverse system size $L_\perp$, compared to local observables, in spite of occurring on short time scales $t\approx 1$. 
This phenomenon, which appears to be at odds with bounds on information propagation, originates from the fact that rate functions are not local observables, but rather exponentially suppressed global quantities, defined for the whole state.
The fidelity density $f$ can indeed be approximated via the local quantities
\begin{align}
f_k \equiv -\frac{1}{k L_\perp} \log\, \langle \psi(t) | P_k | \psi(t) \rangle,
\end{align}
where $P_k$ is a direct product of projectors $P_0=\ket{v} \bra{v}$ onto the local initial state, $\ket{\psi_0}=\otimes_m \ket{v}_m$, applied in a region comprising $k$ consecutive columns, i.e.\ on a total of $k \times L_\perp $ neighboring spins; see Appendix~\ref{app:local} for further details.
The $f_k$ thus defined immediately generalize the local probes recently introduced for one-dimensional systems~\cite{halimeh2020local,Bandyopadhyay2021}.
However, resolving DQPTs requires $k \rightarrow \infty$, so that the $f_k$ become increasingly non-local and the corresponding overlap is exponentially suppressed, besides requiring the simultaneous measurement of an increasing number of spins; this is likely to give rise to experimental complications. 
However, if such experimental issues can be overcome, it might be possible that the exceptional sensitivity of DQPTs to non-local physics might have useful applications. 

\section{Connectivity Effects}\label{sec:connectivity}

\subsection{Interaction-driven pDQPTs}

\begin{figure}[tb]
\begin{center}
\includegraphics[width=0.99\columnwidth]{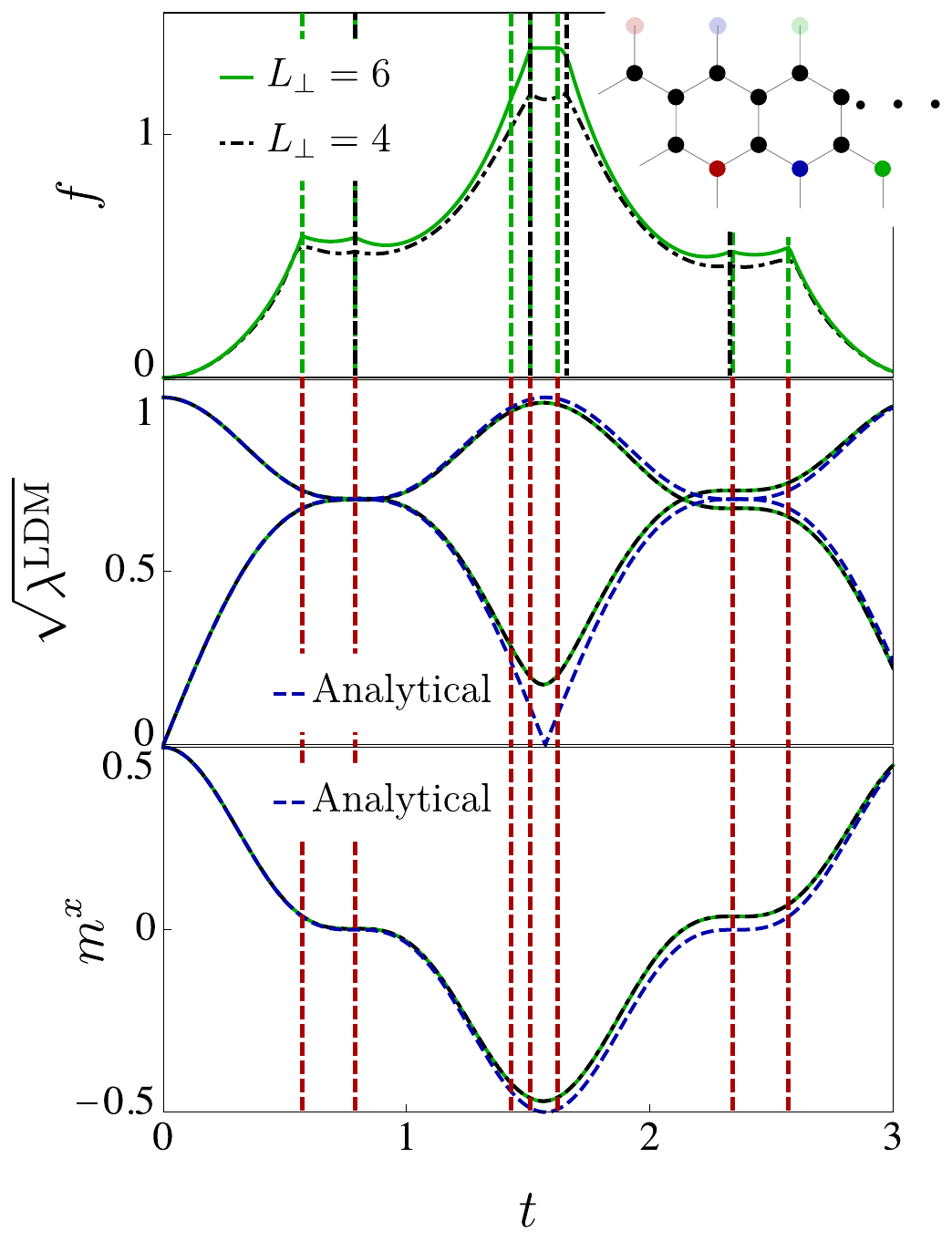}
\end{center}
\caption{\label{fig:c3observables} A quench from the $\ket{\psi_0}=\otimes_m \ket{\rightarrow}_m$ initial state to the strong interaction regime, $h_x=0.1$, $h_z=0$, $J=1$, for a semi-infinite honeycomb lattice reveals a large number of DQPTs. Local probes shown in the bottom panels point to a pDQPT mechanism for crossings happening at $1<t<2$. A comparison between different transverse dimensions $L_\perp=4,6$ reveals the sensitive behavior of the fidelity density. In contrast, the behavior of local observables is accurately captured by the approximate analytical formulae (dashed blue lines) given in the main text and is almost independent of $L_\perp$.  The inset shows an example of the lattice with $L_\perp=4$.}
\end{figure} 

Previously we have found that the predominance of precession-inducing terms, such as strong external fields, typically gives to a certain DQPT phenomenology, characterized e.g. by a precessing behavior of $m(t)$ and a wide entanglement gap. 
On the contrary, strong interactions lead to DQPTs associated with a suppression of local observables and a narrow entanglement gap. 
These mechanisms were found to persist in $D>1$, although the behavior of the resulting DQPTs was found to not always be stable.
However, the possibility of considering different connectivities $c$ for $D>1$ opens up new scenarios, where the key physics can nonetheless be understood in terms of simple underlying mechanisms.

In Fig.~\ref{fig:c3observables} we show one such example by considering a 2D semi-infinite Honeycomb lattice with finite dimension $L_\perp=4$ and $6$ with periodic boundary conditions along the transverse dimension, as shown in the inset. The system is initialized in the $\otimes_m \ket{\rightarrow}_m$ product state and evolved with the Ising Hamiltonian with isotropic interactions $J=1$ and a transverse field $h_x=0.1$ ($h_z=0$). The fidelity density in Fig.~\ref{fig:c3observables}(top) reveals a number of DQPTs. Since the $J$ coupling is dominant in the Hamiltonian, we expect the occurrence of eDQPTs that indeed happens at early time. However,  in addition to the expected eDQPTs, we find a number of DQPTs occurring around $t\approx 1.5$ where the behavior of local probes is characteristic of pDQPTs (as shown by entanglement spectrum and local observables), even though the Hamiltonian is dominated by interactions.

\begin{figure}[tb]
\begin{center}
\includegraphics[width=0.8\columnwidth]{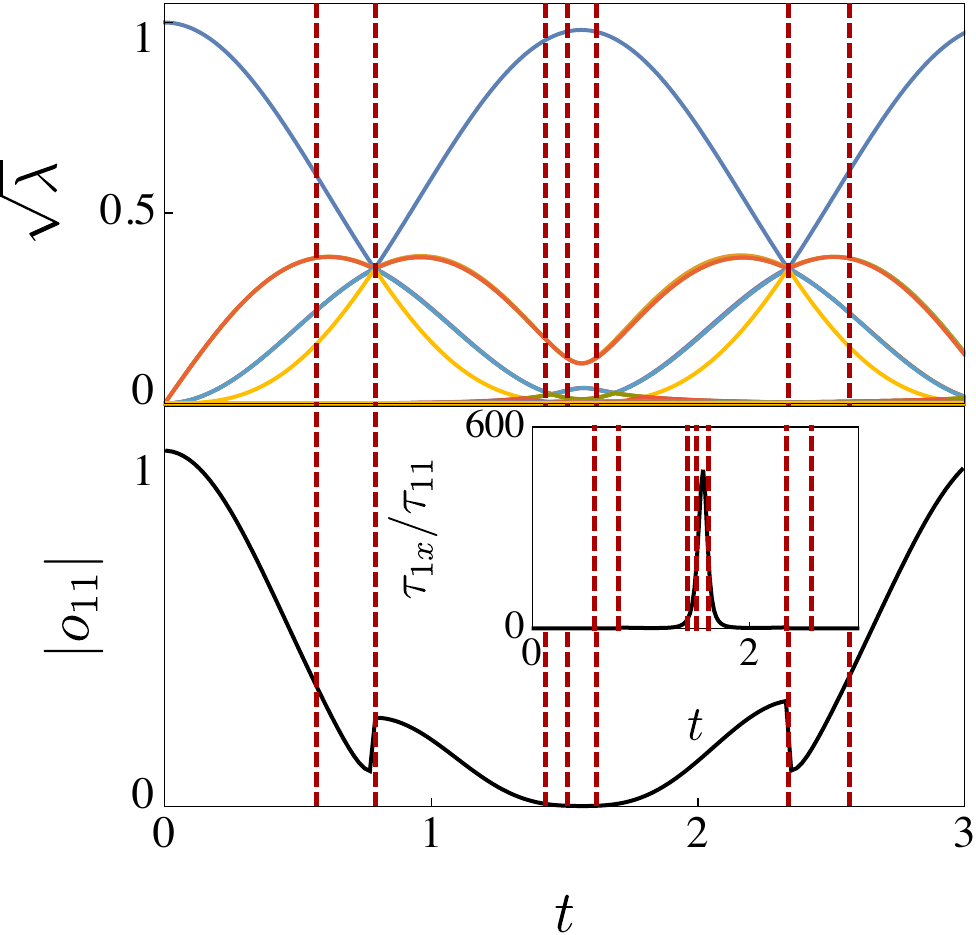}
\end{center}
\caption{\label{fig:c3overlaps} 
For the quench of Fig.~\ref{fig:c3observables} in the honeycomb lattice with $L_\perp=6$, the dynamics of the bipartite entanglement spectrum and leading overlap confirm that both eDQPTs and pDQPTs exist, in spite of the dynamics being largely driven by interactions. Specifically, eDQPTs are associated with entanglement (avoided) crossings, while pDQPTs correspond to a wide entanglement gap and a minimum of $o_{11}$. The occurrence of pDQPTs is also associated with a large relative transition amplitude $\tau_{1x}/\tau_{11}$, as previously found in Fig.~\ref{fig:pDQPToverlaps}. }
\end{figure}

Similarly to the case of a square lattice, the fidelity density $f$ depends sensitively on the transverse size $L_\perp$, as shown by a comparison to $L_\perp=4$. In Fig.~\ref{fig:c3overlaps} we analyze different contributions to the transfer matrix, and observe that both  overlaps and transverse entanglement gap confirm the pDQPT nature of the cusps in the fidelity density happening for $1<t<2$.
Again, we observe that the subleading eigenvalue $\lambda_2$ is nearly threefold-degenerate in that region, with eight-fold degeneracy near the eDQPTs; this potentially explains the appearance of multiple DQPTs as originating from different contributions.
Thus, for the honeycomb lattice, in addition to eDQPTs, we observe pDQPTs that are caused by strong interactions, a possibility that was not observed in one-dimensional spin chains.

\subsection{Analytical description of dynamics}
Below we show that the appearance of pDQPTs in interaction-dominated quenches is in fact a general feature of lattices that have an odd number of nearest neighbors (connectivity), denoted as $c$, which is equal to three for the honeycomb lattice. To this end we analytically compute the local (one-site) reduced density matrix (LDM), $\rho_1$, which gives access to local entanglement and magnetization.
This can be done exactly for arbitrary connectivity $c$ in the classical limit $h_x=0$, since in this case different terms in the Hamiltonian commute and local observables at arbitrary time $t$ can be obtained from a shallow quantum circuit of unit depth. 
In the presence of a small transverse field $h_x$ an approximate equation can still be obtained by a similar method to that used to obtain the analytical eDQPT Ansatz in Ref.~\cite{entanglementView}. 

To compute $\rho_1$, we consider a single spin interacting with $c$ neighbors.
We initialize the system in a product state in the computational basis specified by coefficients $a$ and $b$, $v=(a,b)$. Thus, for spins pointing along the $x$-direction, $a=b=1/\sqrt{2}$. 
By tracing out all spins except for the central one, as shown in Appendix~\ref{app:LDM}, we obtain
\begin{eqnarray}\label{Eq:rho1}
\rho_1 &=& U^\dagger
\begin{pmatrix}|\bar{a}|^2 & \bar{a} \bar{b}^{*} \bar{g}^{c}\\
\bar{a}^{*} \bar{b} (\bar{g}^*)^c & |\bar{b}|^2\end{pmatrix}  U, \\
U &=& e^{i (h_x \sigma^x+ h_z \sigma^{z}) t/2},
\end{eqnarray}
where $\bar{g} = |\bar{a}|^2 e^{-i 2 Jt}+|\bar{b}|^2e^{i 2 Jt}$ and $\bar a,$ $\bar b$ are obtained by acting on the vector by the same matrix $U$ defined above:
\begin{align} \label{Eq:rotate}
\begin{pmatrix} \bar{a} \\ \bar{b} \end{pmatrix} &= U^\dagger \begin{pmatrix} a \\ b \end{pmatrix}.  
\end{align}

From the above expression for the LDM, it is possible to analytically obtain the local entanglement spectrum and local magnetization.
This are in general given by complicated formulae, which can be however simplified in special cases. 
For instance, for $\ket{\psi_0} = \otimes_m \ket{\rightarrow}_m$ 
and $h_z=0$ as in Fig.~\ref{fig:c3observables} the local entanglement and magnetization for general connectivity $c$ are given in closed form by
\begin{align}
\lambda^{\text{LDM}}_{1,2}  &= \frac{1}{2} \left( 1 \pm [\cos(2Jt)]^c \right), \label{eq:lambdageneralc}  \\
m^x & = [\cos(2 J t)]^c .\label{eq:mgeneralc}
\end{align}
Eqs.~\eqref{eq:lambdageneralc} and \eqref{eq:mgeneralc} show that for \emph{odd connectivity} $c$ interactions can induce an effective spin precession, as demonstrated by the wide entanglement gaps, signaling that the state is close to a product state, and the magnetization sign changes, $m^x(t) \approx \pm m^x(0)$, found at times $t =n\pi /2J $, $n \in \mathbb{N} $.

Thus, our findings show that increasing the system's dimensionality opens up new DQPT scenarios compared to one dimension.
In the honeycomb ladder we considered, both pDQPTs and eDQPTs are found in a strong-interaction regime that had been so far associated to eDQPTs only.
However, the precise number and location of DQPTs appears once again to be irregular, making it difficult to extrapolate to the limit of a two-dimensional honeycomb lattice, $L_\perp \rightarrow \infty$.

\section{Conclusion}\label{sec:conclusion}
In this manuscript we investigated the nature of DQPTs on semi-infinite lattices with a finite width considering the quantum Ising model. 
By first studying square lattices, we found that in the strong-field regime one encounters precession-driven DQPTs, previously identified in 1D for similar quench parameters. The pDQPTs on finite width lattices and potentially in two-dimensional systems are still predominantly generated by semiclassical precession, but a complication arises compared to the one-dimensional picture due to the presence of near degeneracies in the entanglement spectrum.
When interactions dominate the dynamics, eDQPTs are still generated by the same mechanism as in 1D.
However, eDQPTs on finite-width lattices are found to be extremely sensitive to the details of the quench and the lattice width.  Here the competing contributions coming from entanglement in the perpendicular and transverse directions lead to complex behavior, effectively paralleling the hybrid regime between p- and eDQPTs previously reported in one dimension~\cite{entanglementView} when both fields and interactions are relevant.

Going beyond square lattices, we also considered the effect of lattices with different number of nearest neighbors (connectivity), $c\neq 4$. 
We found that for lattices with odd-valued connectivity it is possible to observe \textit{interaction-driven} pDQPTs, which show identical phenomenology to pDQPTs but are caused by entanglement-generating terms such as strong Ising two-spin interactions. 
We illustrated this using a particular quench on a honeycomb lattice of finite width with  $c=3$, and also provided a general analytical expressions for general $c$. This suggests that other relatively simple DQPT scenarios beyond those reported in this manuscript might also exist, depending on the details of the system at hand.

In summary, we found that the previously defined p- and eDQPTs generalize to higher dimensional systems represented by lattices of the finite width. For ladders, the same physical mechanisms as in 1D give rise to additional complexities when it comes to DQPTs, including the emergence of eDQPTs purely from interactions. In addition, the extreme sensitivity of DQPTs on the finite width of lattices hinders the extrapolation to a truly thermodynamic limit. While on the one side, this may become an obstacle on the way to observing DQPTs in two-dimensional systems, on the other side such sensitivity could be potentially beneficial for benchmarking unitary evolution algorithms and real quantum simulation devices.

\emph{Acknowledgments.---}
We acknowledge support by the European Research Council (ERC) under the European Union's Horizon 2020 research and innovation programme (grant agreement No.~850899). S.D.N. also acknowledges funding from the Institute of Science and Technology (IST) Austria, and from the European Union's Horizon 2020 research and innovation programme under the Marie Sk\l{}odowska-Curie grant agreement No.~754411.

\appendix 

\section{Approximating DQPT via local projectors}\label{app:local}

\begin{figure}[tb]
\begin{center}
\includegraphics[width=0.99\columnwidth]{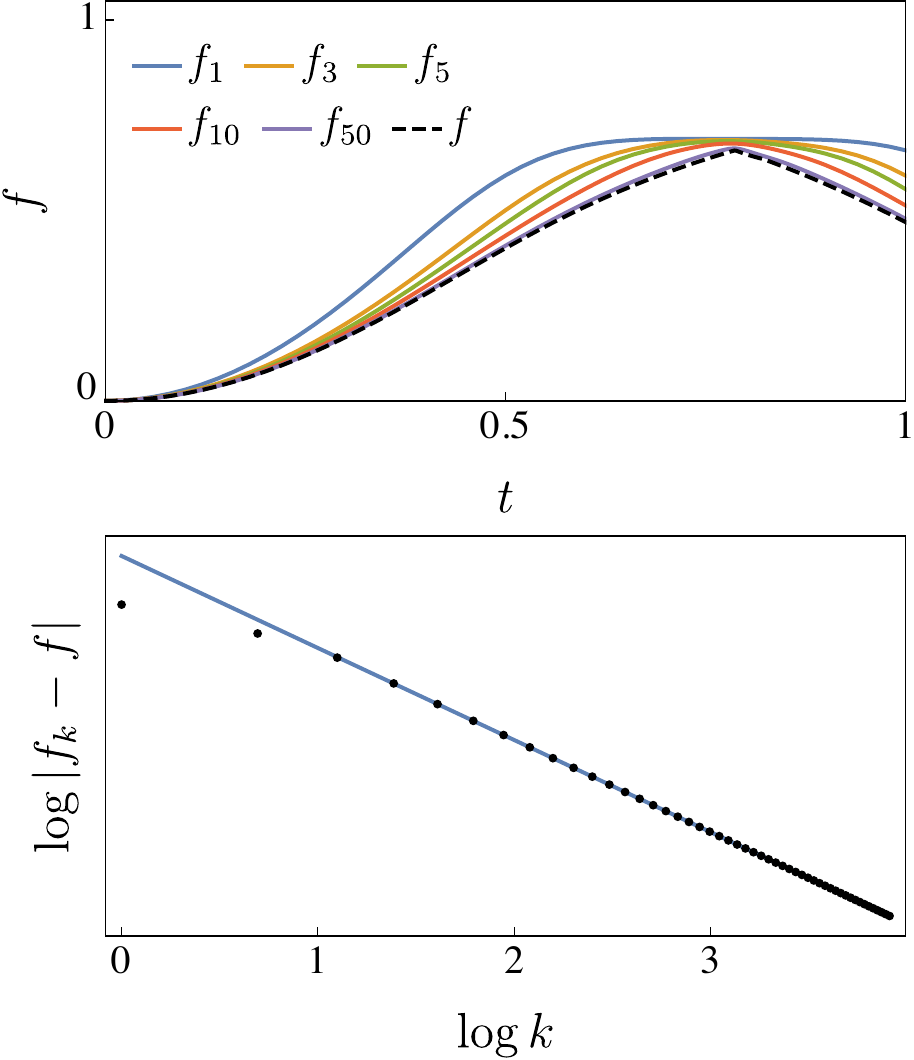}
\end{center}
\caption{\label{fig:fks}
Top panel: fidelity density $f$ and local quantities $f_k$ for different values $k$, for a quench from the $\otimes_m \ket{\rightarrow}_m$ initial state evolved using the Hamiltonian~\eqref{eq:ising} with $J_\parallel=J_\perp=1$, $h_x=0.1$, $L_\perp=3$. As $k$ is increased, the $f_k$ approximate $f$ increasingly well. 
Bottom panel: Power law relaxation of $f_k$ to $f$ as a function of $k$. The full line shows a fit $\log |f_k - f| = a + bk $ with $a\approx -0.84 $, $b\approx -1$. 
The results shown in this plot are obtained using iTEBD as explained in the main text. }
\end{figure}

For an initial product state, $\ket{\psi_0} =\otimes_m\ket{v}_m$, DQPTs in the semi-infinite geometry of this manuscript can be experimentally probed by measuring suitable combinations of the local projectors onto the initial state, $P^0_m= \ket{v}_m \bra{v}_m$, which generalize the local quantities introduced in 1D~\cite{halimeh2020local,Bandyopadhyay2021}. 
Namely, one considers the local quantities 
\begin{align}
f_k \equiv -\frac{1}{k L_\perp} \log\, \langle \psi(t) | P_k | \psi(t) \rangle ,
\end{align}
where $P_k=\otimes_{m\in \mathcal{S}_k} P^0_m$ and $\mathcal{S}_k$ is a region comprising $k$ consecutive columns, each of length $L_\perp$.
As $k$ is increased, $f_k$ increasingly well approximates $f$, as shown in Fig.~\ref{fig:fks}. We note that $f_{k}$ are closely related to the local order parameters, such as the magnetization, that are commonly employed to estimate the transition point, e.g. for $P_{1} = \ket{\uparrow}\bra{\uparrow} = \frac{1 + \sigma^z}{2}$. In this manuscript, we stick to traditional local observables instead of $f_k$.

\section{Computational details}\label{app:computational}
In this manuscript, we represent a semi-infinite ladder as an iMPS where each site represents a column of length $L_\perp$. The system can then be time-evolved used iTEBD~\cite{Vidal2006}. 
For the present system the local physical dimension is however $d=2^{L_\perp}$, rather than $d=2$ as common in iTEBD applications.
Thus, in order to improve computational efficiency, we include a number of additional steps, outlined below, over the standard iTEBD algorithm.

We begin by performing a second-order Trotter decomposition, whereby for a chosen time step $\Delta t$ the time-evolution operator is approximately decomposed into a product of $n = t/\Delta t$ operators:
\begin{align}
U(t) = e^{-iHt} \approx[ U_A(\Delta t/2) U_B(\Delta t) U_A(\Delta t/2)]^n ,
\end{align}
where $U_A = \prod_{i  \in \text{even}} U^{(i)}$, $U_B = \prod_{i \in \text{odd}} U^{(i)}$ and $U^{(i)}$ only acts on the two consecutive sites $(i,i+1)$~\cite{Vidal2006}, which here correspond to two columns.
For each $U^{(i)}$, we now perform an additional second-order Trotter decomposition
\begin{align}\label{eq:trotter2}
U^{(i)}(t) \approx U^{(i)}_L(t/2) U^{(i)}_I(t) U^{(i)}_L(t/2) 
\end{align}  
to separate the contribution of local fields, included in $U^{(i)}_L$, from the interactions within and across the two columns, included in $U^{(i)}_I$.
For the model~\eqref{eq:ising}, interactions are diagonal in the $z$-basis, so that $U^{(i)}_I$ is a diagonal matrix and its application on the local state can be performed as the element-wise multiplication of two vectors. 
This replaces an operation whose computational cost scales as $\chi^2 2^{4L_\perp}$ with one scaling as $\chi^2 2^{2L_\perp}$. 

A further speed-up is achieved by replacing the subsequent SVD by a reduced-rank randomized singular value decomposition (RRSVD)~\cite{rSVD2015}. 
This replaces the cost of directly performing the SVD of a $\chi  2^{L_\perp} \times \chi  2^{L_\perp} $ matrix, scaling as $\chi^3  2^{3L_\perp}$, with a number of operations whose complexity scales at worst as $\chi^3 2^{2L_\perp}$.
In our application of RRSVD, we fix the final bond dimension $\chi$ upfront (there also exists an algorithm to dynamically adjust the bond dimension based on a target accuracy, which however would here entail an additional computational cost~\cite{rSVD2015}).
In one dimension, it was found that, on the time scales of interest for early-dynamics p- and eDQPTs, most of the quantum dynamics is encoded in the two leading singular values; this observation lies at the root of the $\chi=2$ analytical DQPT Ans\"atze~\cite{entanglementView}.
In this manuscript, we consider systems made up of $L_\perp$ coupled chains and quenches and time-scales that are very similar to those of Ref.~\cite{entanglementView}.
Based on the behavior in 1D, it would then be reasonable to expect that in this regime the required bond dimension be of order $\chi\approx 2^{L_\perp}$.
Empirically, considering different values of $L_\perp$, we indeed observe a sharp drop in the magnitude of $\lambda_i$ for $i>2^{L_\perp}$.
For each quench, by extrapolating from smaller system sizes, we checked that increasing the bond dimension beyond $\chi=2^{2_{L_\perp}+1}$ does not affect our results, including the observables, the fidelity, the overlaps or the singular values (namely, although there are more $\lambda_i$ for larger $\chi$, the leading ones still match and the additional ones take very small values). In all cases, we found that for $\chi=2^{2_{L_\perp}+1}$ the maximum error in the fidelity density $f$ is below $1 \%$ (i.e. not visible on the scale of the present plots).
Thus, we set the bond dimension to $\chi=64>2^{L_\perp+1}$ for $L_\perp = 3,4$ and $\chi=2^{L_\perp+1}$ for $L_\perp=5,6$.
We use a time step of $\Delta t=0.01$, with smaller time steps used in the vicinity of DQPTs to achieve better resolution. We checked that further decreasing the time step does not lead to appreciable changes.

\section{Two-dimensional analytical Ans\"atze}\label{app:ansatz}
\begin{figure}[tb]
\begin{center}
\includegraphics[width=0.95\columnwidth]{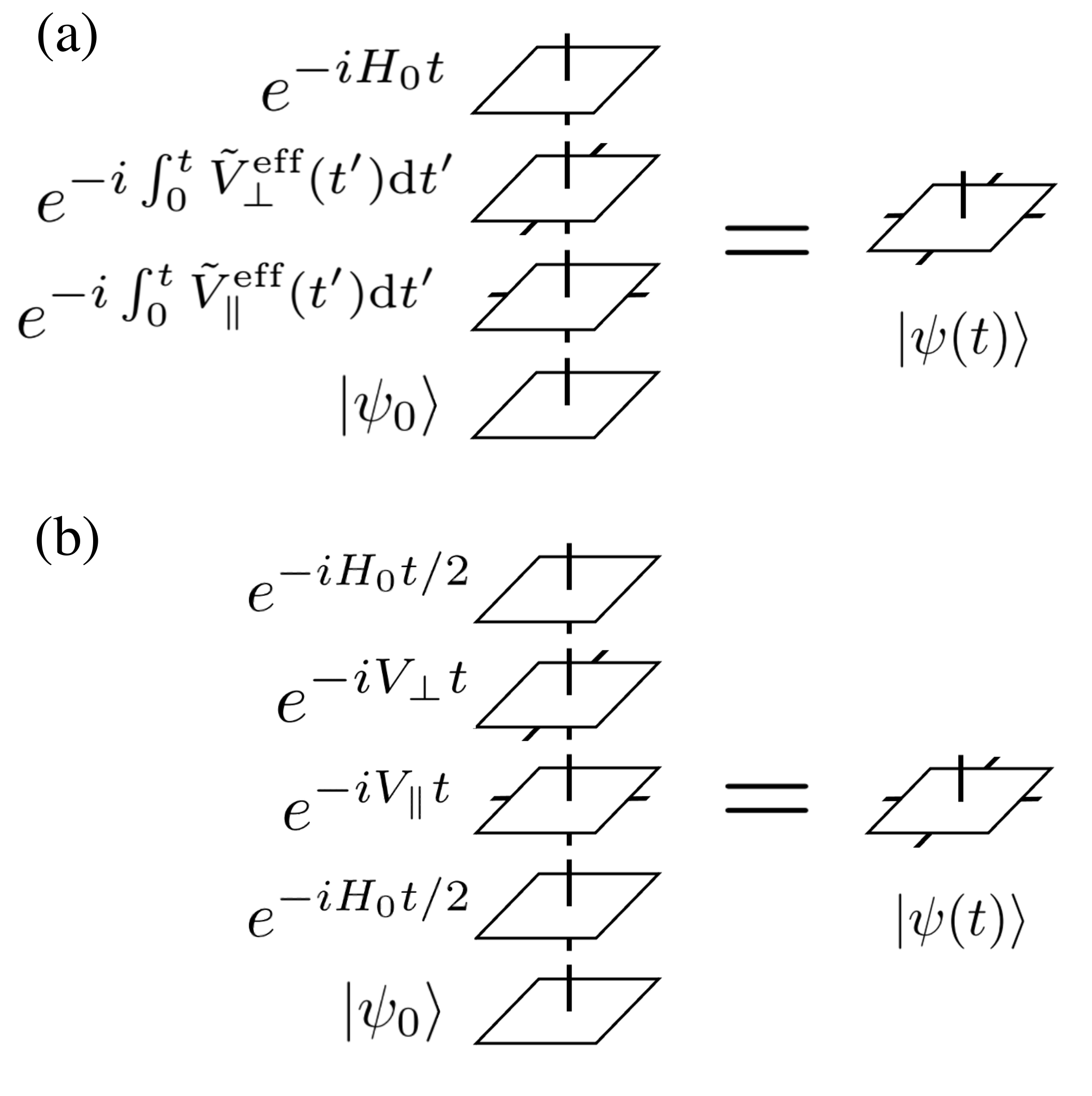}
\end{center}
\caption{\label{fig:ansatz} 
Construction of the analytical PEPS Ans\"atze for (a) pDQPTs and (b) eDQPTs. }
\end{figure}
In this Appendix, we provide a derivation of two analytical Ans\"atze generalizing those introduced in Ref.~\cite{entanglementView}, which were designed to capture the short-range physics leading to p- and eDQPTs.
Their two-dimensional generalizations provide an extra handle to assess to what extent two-dimensional DQPTs can be ascribed to the same mechanisms as in 1D.
While in one dimension the Ants\"atze take the form of $\chi=2$ iMPS, in two dimensions the construction naturally produces a PEPS with physical dimension $d=2$ and uniform bond dimension $\chi=2$.
The generalization is straightforward and the resulting PEPS can be written out analytically; however, due to the complexity of contracting two-dimensional lattices, the PEPS Ans\"atze do not immediately yield closed-form expressions unlike their MPS precursors.
The PEPS obtained from the Ans\"atze must thus be contracted numerically in order to calculate physical quantities. 
However, the PEPS represents the state at a general time $t$, without needing to time-evolve numerically; this avoids the computational bottlenecks associated with time evolution (see Appendix~\ref{app:computational}).

\subsection{pDQPT Ansatz}
\begin{figure}[tb]
\begin{center}
\includegraphics[width=0.8\columnwidth]{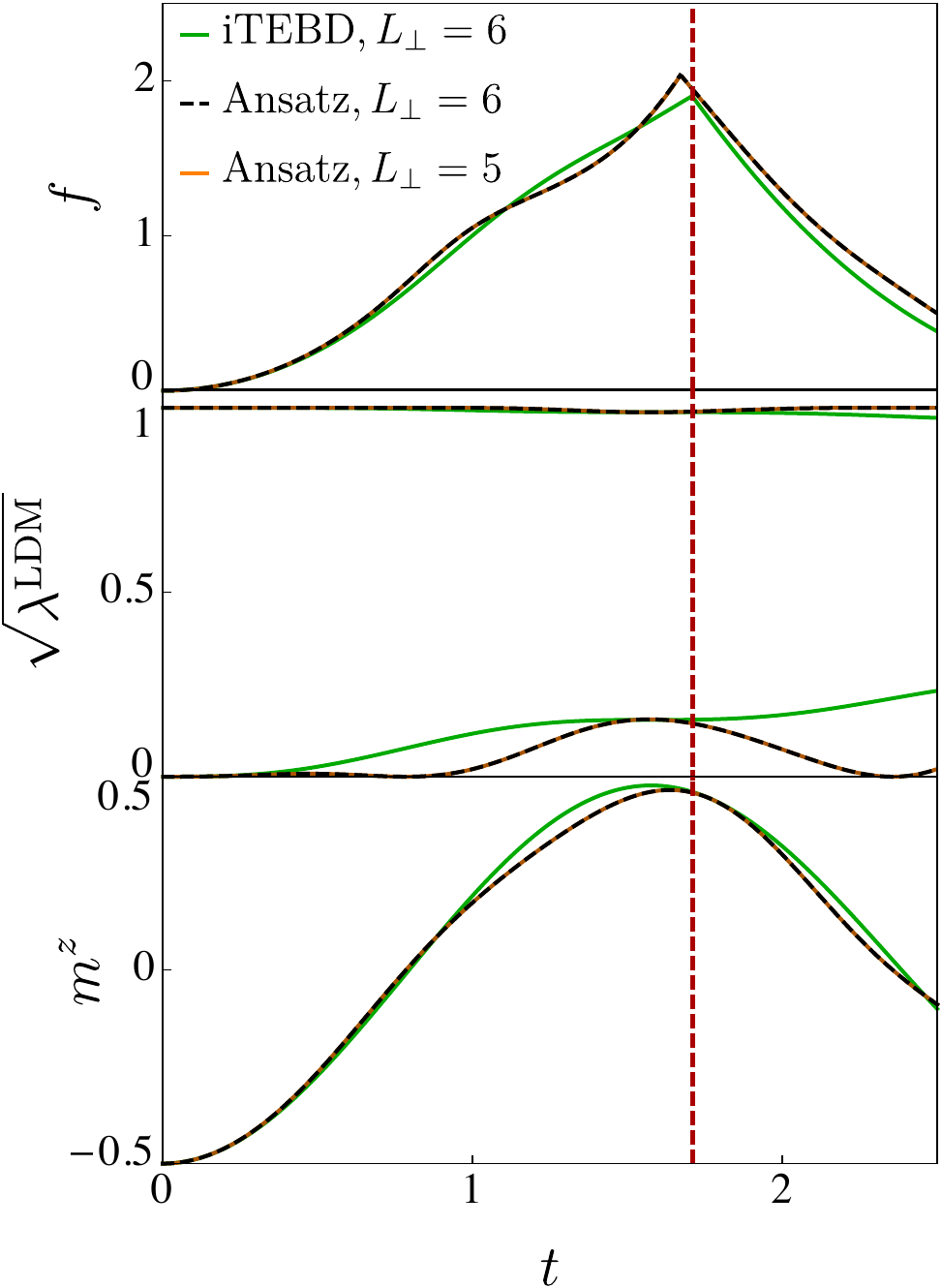}
\end{center}
\caption{\label{fig:pDQPtansatz} 
PEPS pDQPT Ansatz describing the 2D quantum Ising model on a square lattice in the strong field regime. For the quench of Fig.~\ref{fig:pDQPT}, the Ansatz captures the correct qualitative dynamics of the fidelity, local entanglement and local magnetization corresponding to a pDQPT and shows reasonable quantitative agreement over the considered time range. However, it does not capture the disappearance of the peak for $L_\perp=5$ observed in Fig.~\ref{fig:pDQPT}. }
\end{figure}

\begin{figure}[tb]
\begin{center}
\includegraphics[width=0.8\columnwidth]{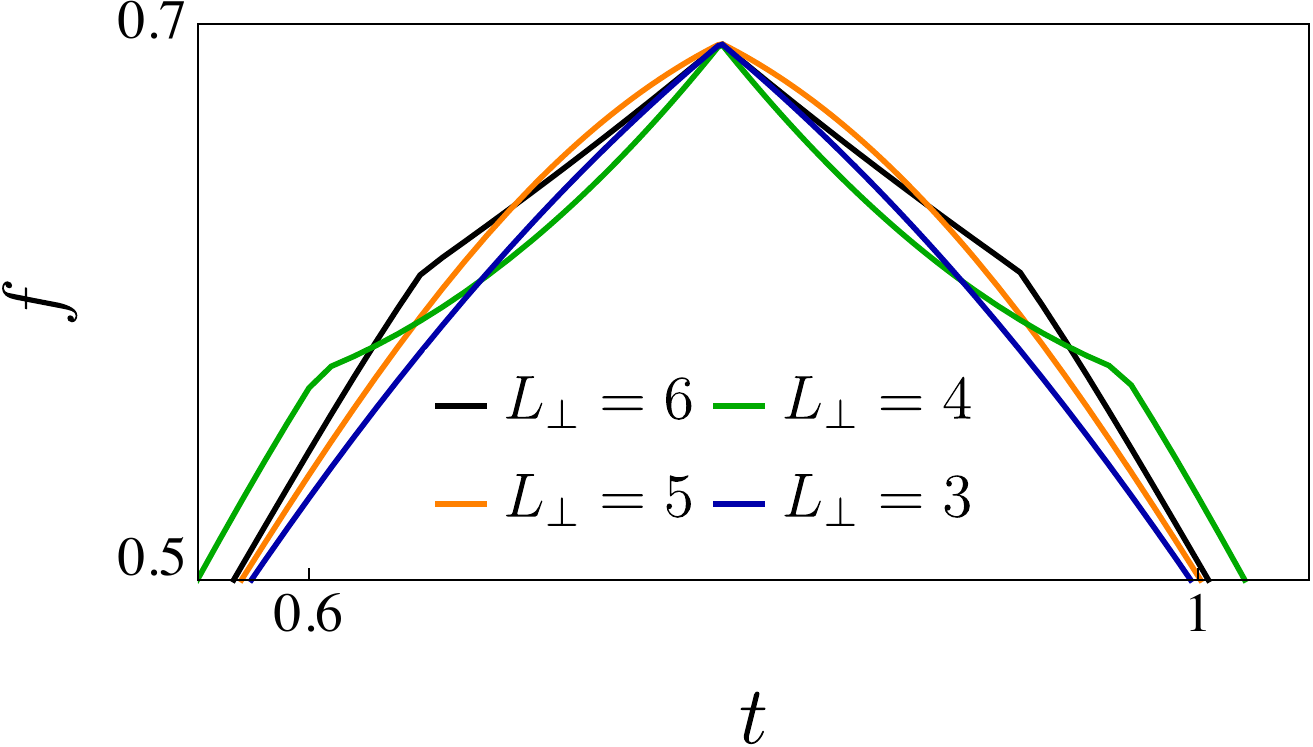}
\end{center}
\caption{\label{fig:eDQPtansatzTop} 
Detail of the first peak region for the analytical eDQPT ansatz, describing the Ising model on a semi-infinite square lattice in the strong-interaction regime, for different transverse system sizes $L_\perp$. We consider the quench of Fig.~\ref{fig:eDQPT}. It can be seen that, in spite of agreeing on the central peak, different $L_\perp$ correspond to different positions and numbers of DQPTs, indicating slow convergence with $L_\perp$. In particular, we observe indications of an odd/even effect, with a single DQPT arising for odd $L_\perp$ whereas three are present for even $L_\perp$. }
\end{figure}

Let us begin with the analytical pDQPT Ansatz, which is constructed to capture the dynamics in the limit $h_x, h_z \gg J_\parallel, J_\perp $.
The Hamiltonian~(\ref{eq:ising}) can be separated into a free-precessing part containing only single-spin terms,  $H_0= \sum_{ij}[h_{x}\sigma_{ij}^{x}+h_{z}\sigma_{ij}^{z}]$, and an interacting part, which we further split into a parallel and a transverse component, $V=V_\parallel + V_\perp = \sum_{ij} [ J_\parallel \sigma^z_{i j} \sigma^z_{i+1 j} + J_\perp  \sigma^z_{i j} \sigma^z_{i j+1} ]$.
Following Ref.~\cite{entanglementView}, we move to the rotating frame with respect to $H_0$:
\begin{align}
\ket{\psi(t)}  = e^{-i  H_0 t} \mathrm{T} e^{-i\int_0^t {\tilde V}(t^\prime)  \mathrm{d}t^\prime} \ket{\psi_0}.
\end{align}
The rotating frame operators $\tilde{V} = \tilde{V}_\parallel + \tilde{V}_\perp$ are straightforwardly obtained by summing, respectively over columns and rows, copies of the operator $\tilde{V}_{\text{1D}}$ obtained from the one-dimensional operator $V_{\text{1D}} = J \sum_i \sigma^z_i \sigma^z_{i+1}$:
\begin{align}
{\tilde V}_{\text{1D}}(t) = e^{i t H_0} V e^{-i t H_0} \equiv  \sum_{i}\sum_{\alpha,\beta} s_\alpha(t) s_\beta(t) \sigma^\alpha_i\sigma^\beta_{i+1}
\end{align}
where $\alpha, \beta \in \{x,y,z\}$ and we defined $s_x(t) = 2 h_xh_z\sin ^2(h t)/h^2$, $s_y(t)= h_x\sin (2 h t) /h$, $s_z(t) =[ h_x^2 \cos (2 h t)+h_z^2]/h^2$, $h = \sqrt{h_x^2+h_z^2} $.
By further approximating the $x$ and $z$ operators by their expectation values under free precession~\cite{entanglementView}, $\sigma^x  \rightarrow -s_x$ and $\sigma^z\rightarrow -s_z$, $\tilde{V}_{\text{1D}}$ can be expresses in terms of y operators only:
\begin{align}
\tilde{V}_{\text{1D}}(t) \approx \tilde{V}_{\text{1D}}^{\text{eff}}(t) = \sum_i [  J^{\text{eff}}(t) \sigma_i^y \sigma_{i+1}^y + h^{\text{eff}}(t) \sigma_i^y]
\end{align}
with $J^{\text{eff}} = J  s_y^2, \quad h^{\text{eff}} = -2 J s_y (s_x^2+ s_z^2 ) $.
With this approximation, the interaction term can be straightforwardly exponentiated as a $\chi=2$ matrix product operator (MPO)~\cite{entanglementView}, $\mathrm{T} e^{-i\int_0^t {\tilde V}(t^\prime)  \mathrm{d}t^\prime} \approx \prod_i U_i$ with
\begin{align}
U_i =\left(\begin{matrix}e^{-i J a(t) - i J b(t) }P^y_i & e^{i J a(t)  - i J b(t) }P^y_i \\ e^{iJ a( t) + i J b(t)  } P^{-y}_i  & e^{-i J a(t) +  i J b(t) } P^{-y}_i \end{matrix} \right) ,
\end{align}
where $P_i^{\pm y }\equiv \ket{\pm y}_i \bra{\pm y}_i$ are the projectors on the $y$-eigenstates, $\sigma^y \ket{\pm y} = \pm \ket{\pm y}$. 
The effective interactions $\tilde{V}^{\text{eff}}_\parallel$, $\tilde{V}^{\text{eff}}_\perp$ are obtained by following the same steps as discussed above. $\tilde{V}^{\text{eff}}_\parallel$ and $\tilde{V}^{\text{eff}}_\perp$ commute, so that for the two-dimensional case we can write
\begin{align}
\mathrm{T} e^{-i\int_0^t {\tilde V}(t^\prime)  \mathrm{d}t^\prime} \approx e^{-i\int_0^t {\tilde V^{\text{eff}}_\parallel}(t^\prime)  \mathrm{d}t^\prime} e^{-i\int_0^t {\tilde V^{\text{eff}}_\perp}(t^\prime)  \mathrm{d}t^\prime}.
\end{align}
From the one-dimensional case, we know that the exponentials featuring $\tilde V_\perp(t)$ and $\tilde V_\parallel(t)$ can both be written as parallel copies of $\chi=2$ MPOs, respectively representing columns and rows.
An approximation to the full time-evolution operator is then obtained by a subsequent application of these MPOs, which gives rise to a rank-$6$ tensor; see Fig.~\ref{fig:ansatz}.
Acting by this operator on the initial state gives the PEPS ansatz for $\ket{\psi(t)}$.

\begin{figure}[tb]
\begin{center}
\includegraphics[width=0.8\columnwidth]{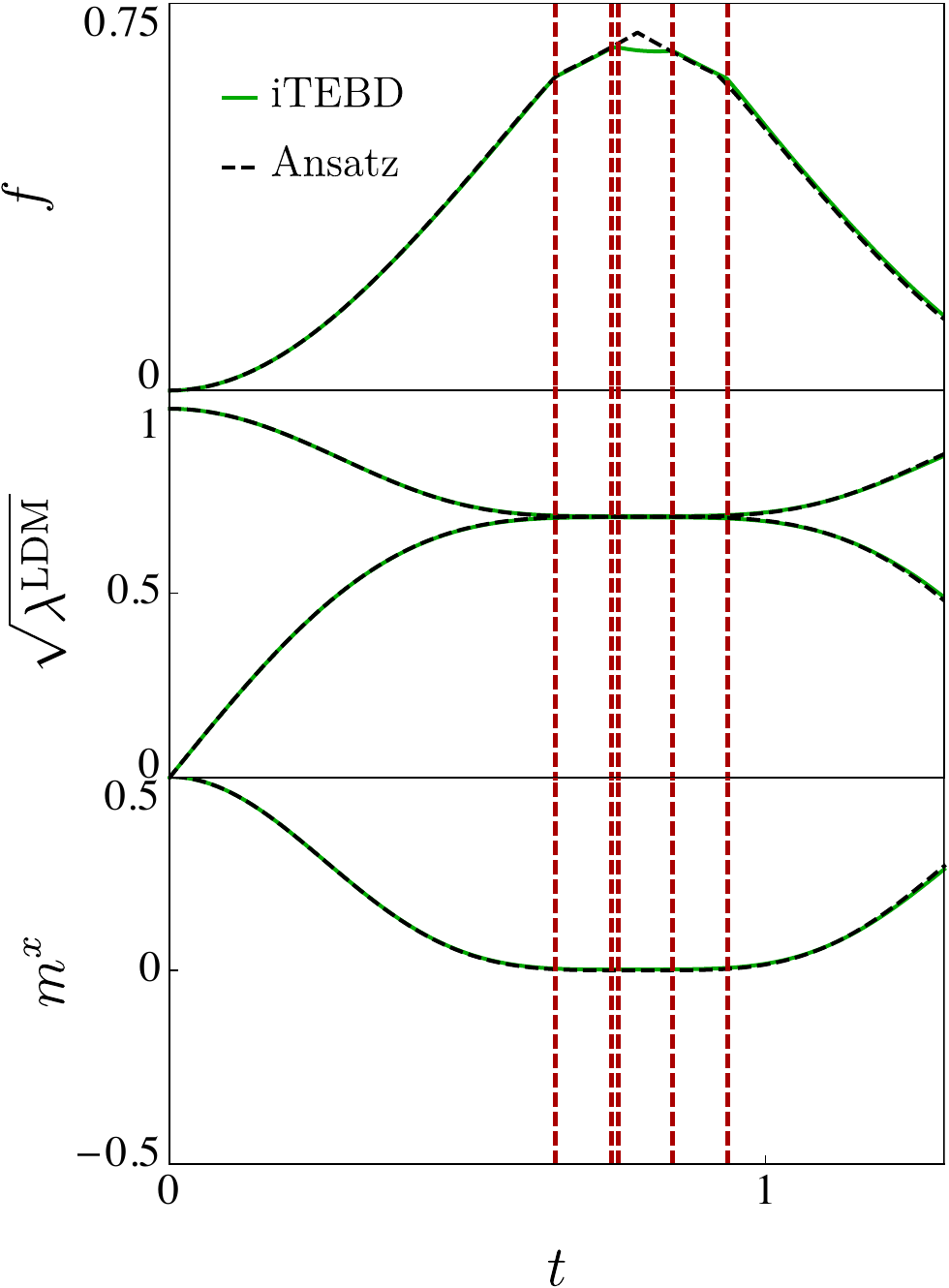}
\end{center}
\caption{\label{fig:eDQPtansatz} 
Comparison of the eDQPT ansatz and iTEBD for the quench of Fig.~\ref{fig:eDQPT} and $L=6$. The vertical dashed lines show the location of DQPTs obtained from the iTEBD result. The ansatz largely agrees with iTEBD on the location of the peak, but fails to predict the specific details (e.g.\ the exact number of DQPTs observed), with the disagreement being more pronounced in the central region of the picture. This is in spite of nearly perfect agreement in describing the local entanglement spectrum and $x$-magnetization.}
\end{figure}

\subsection{eDQPT Ansatz}
Following Ref.~\cite{entanglementView}, in order to construct an analytical $\chi = 2$ eDQPT Ansatz we again split the Hamiltonian into a single-spin and a two-spin term, $H_0$ and $V=V_\parallel + V_\perp$, and approximate the time-evolution operator by a second-order Trotter decomposition:
\begin{align}\label{eq:eDQPTansatz}
e^{-iHt} \approx e^{-iH_0 t/2}e^{-iV_\parallel t}e^{-iV_\perp t} e^{-iH_0 t/2} ,
\end{align}
where we exploited the commutativity of $V_\parallel$ and $V_\perp$.
Each exponential featuring an interaction term is diagonal in the z-basis and thus admits an exact MPO representation with $\chi=2$, as shown above. Again, as shown in Fig.~\ref{fig:ansatz}, this amounts to stacking two copies of the 1D interaction term at each site, one corresponding to the rows and one to the columns. The PEPS ansatz is then obtained by contracting with the initial state. 
Although the PEPS Ansatz does not immediately yields results in close form, due to the complexity of contracting a higher-dimensional lattice, one-site quantities (such as local observables or the local entanglement) in the same approximation can be computed from the local density matrix discussed in Appendix~\ref{app:LDM}.

\section{Restoring the pDQPT for $L_\perp=5$}\label{app:pDQPT}

\begin{figure}
\begin{center}
\includegraphics[width=0.8\columnwidth]{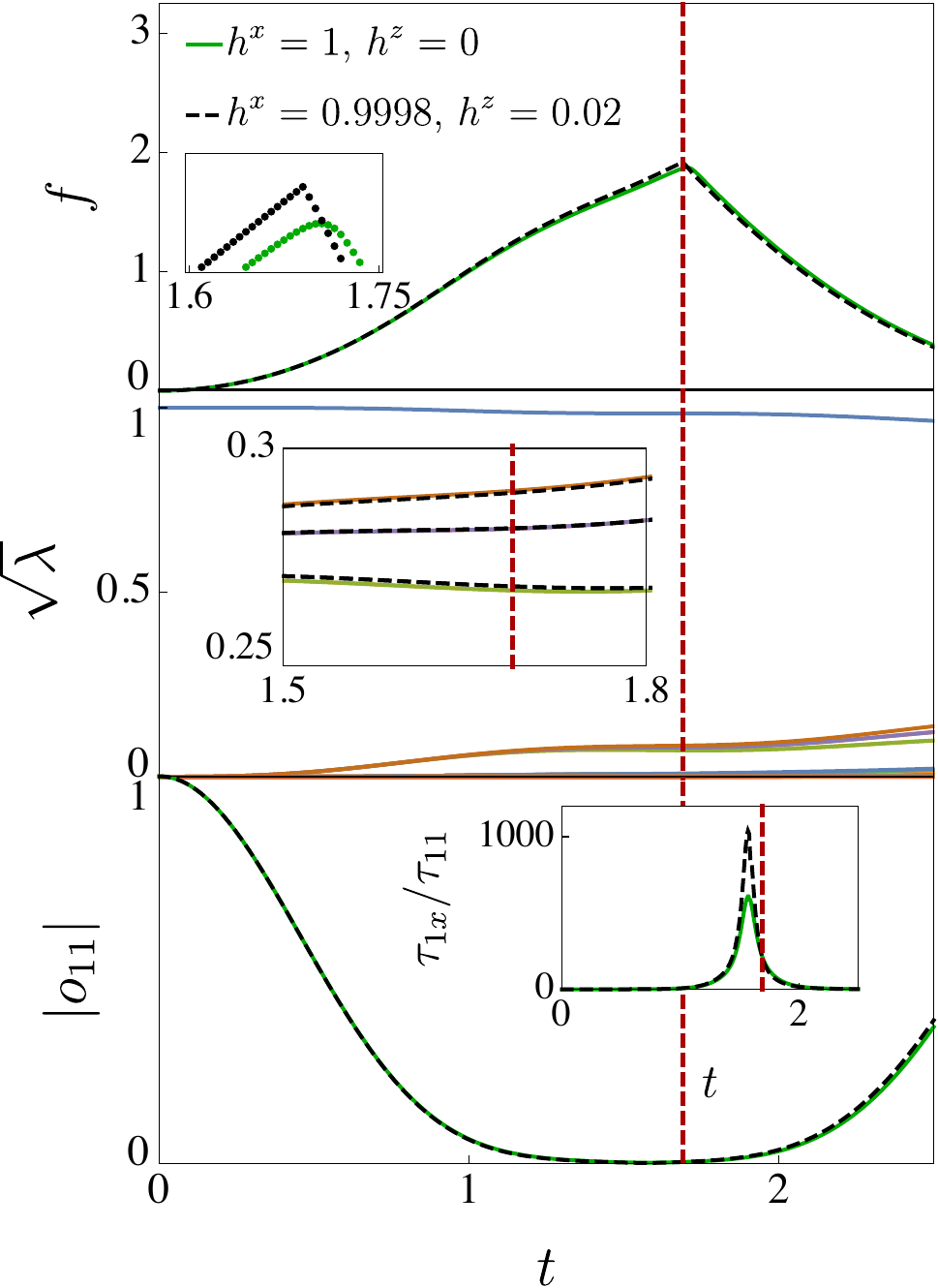}
\end{center}
\caption{\label{fig:pDQPToverlapsL5} 
Fidelity density, column-wise entanglement spectrum $\{ \lambda_i \}$ and overlaps $o_{ij}$ as defined in the main text for the semi-infinite 2D Ising model with $L_\perp=5$ following a quantum quench from the $\ket{\rightarrow}$ product state. We consider $J_\perp=J_\parallel=0.1$ and either $h_x=1$ (solid lines, corresponding to
the quench of Fig.~\ref{fig:pDQPT}) or $h_x=0.9998$, $h_z=0.02$ (dashed lines).
The top panel shows that the DQPT absent  Fig.~\ref{fig:pDQPT} for $L_\perp=5$ can be restored by the inclusion of the small longitudinal $h_z$ field; this is in spite of no significant change occurring in the local physics.
In the middle panel, we observe a large gap between $\lambda_1$ and a small set of singular values $\{ \lambda_i \}$, which are in turn significantly larger than the bulk of the spectrum. 
Again, the inclusion of the longitudinal field does not lead to a qualitative difference, as further highlighted by the detail in the inset.
The bottom panel shows the overlap $|o_{11}|$ corresponding to the dominant product state. The DQPT occurs in the vicinity of the minimum of $|o_{11}|$. As shown in the inset, this is also near a maximum of the overall relative excitation amplitude $\tau_{1x}/\tau_{11}$. In both cases, there is no qualitative difference between the two quenches, however a DQPT is only observed for the quench with non-zero $h_z$.
}
\end{figure}

For the quench of Fig.~\ref{fig:pDQPT}, we saw that no DQPT was observed for $L_\perp=5$, in spite of it occurring for $L_\perp=3,4,6$ and good convergence of the local observables.
To show that this absence of DQPT is likely result of an accidental symmetry or some other fine-tuning, in Fig.~\ref{fig:pDQPToverlapsL5} we include a small perturbation to the above quench, setting the external field to $h_z=0.02$, $h_x=0.9998$ so that the total applied field $h=\sqrt{ (h_z)^2+(h_x)^2}\approx 1$ as before. In the top panel, this perturbation is observed to restore the DQPT observed for other system sizes. The entanglement and overlaps driving the DQPT show very little change upon the rotation of the field. Also local observables (not shown) are nearly indistinguishable for the original and perturbed quenches.

\section{Calculation of the LDM}\label{app:LDM}
In this section we derive the analytical expressions for the one-site reduced density matrix, or local density matrix (LDM), used in the main text.
We first compute the exact result for the classical quenches $J, h_{z}\neq 0, h_x=0$, and then approximately generalize the results to quantum quenches with $h_{x} \neq 0$.

Due to the shallow quantum circuit structure of the classical quench $h_x=0$, we can ignore sites that are not nearest neighbors with the site of interest as they factor out in the calculation. 
We thus consider a spin coupled to $c$ neighbors which are not coupled to each other. The number of neighbors $c$ corresponds here to the connectivity of the lattice. We denote the central spin by $0$ and the remaining ones by $1, \dots, c$.
To calculate the LDM we write the wavefunction as a $2\times 2^c$ matrix $\ket{\psi} = \sum_{i j} C_{i j} \ket{i} \otimes \ket{j}$, where the indices $i,j$ run over $i \in \{ \uparrow, \downarrow\}$ and $j \in \{0,2^c-1\}$, and the integers $j$ denote different configurations of the $c$ spins, e.g. $\ket{0} = \ket{\downarrow\ldots \downarrow}$. 
We choose the initial state to be a product state and permutation invariant (i.e. each site is initialized in the same local state), $\ket{\psi_0} = \otimes_{m=1}^c \ket{\psi}_m$, $\ket{\psi}_m = a\ket{\uparrow}_m + b\ket{\downarrow}_m$ with $|a|^2 + |b|^2 = 1$. 
The initial state can then be written as 
\begin{equation}
\psi(t) = a^{c+1}\begin{pmatrix} 1 & \kappa &\ldots &  \kappa^c \\
 \kappa &   \kappa^2&\ldots &  \kappa^{c+1} \end{pmatrix},
\end{equation}
where $ \kappa = b/a$ and the first row corresponds to the $\ket{\uparrow}$-state of the $0$-th spin, the second row corresponds to the $\ket{\downarrow}$-state, and the columns correspond to the $\ket{j}$.

We begin by considering $h_x,h_{z} = 0$.
The application of the gate $U_{J} = \prod_{j=1}^{c}e^{-i J \sigma^{z}_{0}\sigma^{z}_{j} t}$ on $\ket{\psi_0}$ can be directly evaluated element by element.
The evolved state $\ket{\psi(t)} = U_J \ket{\psi_0}$ is then given by
\begin{equation}
 \psi(t) =
 a^{c+1}
 \begin{pmatrix} e^{-i c Jt} & e^{-i(c-2)Jt}  \kappa &\ldots & e^{i c Jt}  \kappa^c \\
e^{i c Jt} \kappa &  e^{i(c-2)Jt}  \kappa^2&\ldots & e^{-i c Jt} \kappa^{c+1} \end{pmatrix}.
\end{equation}
We now trace out all spins except for the central one.
The diagonal elements of the reduced density matrix $\rho_1 = \psi\cdot \psi^{\dag}$ are trivial since the phases cancel out. The off-diagonal elements can be calculated by noticing that the multiplicity of the various phases is given by binomial coefficients. For example, if the central spin is $\ket{\uparrow}$, the phase $e^{-i(c-2n)J}$ will appear $\binom{c}{n}$ times. 
If the central spin is $\ket{\downarrow}$ the same holds but the corresponding phases are multiplied by a minus sign. 
This makes it possible to re-sum the terms corresponding to the off-diagonal elements as
\begin{align}
\begin{split}
& a b^* |a|^{2c}  e^{-2icJt}  +  \binom{c}{1} a b^* |a|^{2c-1} |b|^2 e^{-2i(c-1)Jt}  \\ 
&+ \binom{c}{2} a b^* |a|^{2c-2} |b|^4 e^{-2i(c-2)Jt} +  \dots  = a b^* g^c
\end{split}
\end{align}
with
\begin{align}
g = |a|^2 e^{-i 2 Jt}+|b|^2e^{i 2 Jt} .
\end{align}
The LDM $\rho_1 = \psi\cdot \psi^{\dag}$ is then given by
\begin{equation}\label{Eq:DM}
\rho_1 = \begin{pmatrix}|a|^2 & a b^{*} g^{c}\\  
a^{*} b (g^*)^c & |b|^2\end{pmatrix} .
\end{equation}
Local fields can be included by approximating the time-evolution operator by a second-order Trotter decomposition $U_{J,h_{x},h_{z}}(t) =  U_{h_{x},h_z}(t/2) U_{J}(t) U_{h_{x},h_z}(t/2)$, where
\begin{align}
\begin{split}
U_{h_x,h_z}(t) &=  \prod_{j=0}^{c} e^{-i (h_x \sigma^x_{j}+ h_z \sigma^{z}_{j}) t}\\
 &= \otimes_{j=0}^{c}  \begin{pmatrix} A(t) & B(t) \\  
B(t) & A^*(t) \end{pmatrix}
\end{split}
\end{align}
and using the magnitude of the field $h = \sqrt{h_x^2 + h_z^2}$ we obtain:
\begin{equation}
A(t) = \cos ( h t) - i \frac{h_z}{h} \sin (h t), \
B(t) = - i \frac{h_x}{h} \sin ( h t). 
\end{equation}
This procedure is exact for $h_z \neq 0$, $h_x=0$ and approximate for non vanishing transverse field $h_x \neq 0$.
To calculate the time-evolved state, we first apply the local unitary gate $U_{\cos h,\sin h}(t/2)$ to the initial state, which amounts to a rotation at each site as defined in Eq.~(\ref{Eq:rotate}) in the main text.
We can then proceed to applying the $U_J$ gate as for the $h_{x,z}=0$ case, which results in a density matrix of the form of Eq.~(\ref{Eq:DM}) with the replacements $a\rightarrow \bar{a}$, $b\rightarrow \bar{b}$. 
Finally, we apply the second local gate which amounts to a further local rotation, arriving to Eq.~(\ref{Eq:rho1}) in the main text. 
This expression is exact for $h_x=0$ and holds approximately for small $h_x$.
From the LDM one can compute the local entanglement spectrum, given by its eigenvalues, and local expectation values $\langle o \rangle$ as $\Tr (\rho_1 o )$.
In fact, the approximation used to compute the LDM is the same one that underpins the two-dimensional eDQPT ansatz discussed in Appendix~\ref{app:ansatz}, so that local quantities are the same in both cases.


\begin{thebibliography}{71}%
\makeatletter
\providecommand \@ifxundefined [1]{%
 \@ifx{#1\undefined}
}%
\providecommand \@ifnum [1]{%
 \ifnum #1\expandafter \@firstoftwo
 \else \expandafter \@secondoftwo
 \fi
}%
\providecommand \@ifx [1]{%
 \ifx #1\expandafter \@firstoftwo
 \else \expandafter \@secondoftwo
 \fi
}%
\providecommand \natexlab [1]{#1}%
\providecommand \enquote  [1]{``#1''}%
\providecommand \bibnamefont  [1]{#1}%
\providecommand \bibfnamefont [1]{#1}%
\providecommand \citenamefont [1]{#1}%
\providecommand \href@noop [0]{\@secondoftwo}%
\providecommand \href [0]{\begingroup \@sanitize@url \@href}%
\providecommand \@href[1]{\@@startlink{#1}\@@href}%
\providecommand \@@href[1]{\endgroup#1\@@endlink}%
\providecommand \@sanitize@url [0]{\catcode `\\12\catcode `\$12\catcode
  `\&12\catcode `\#12\catcode `\^12\catcode `\_12\catcode `\%12\relax}%
\providecommand \@@startlink[1]{}%
\providecommand \@@endlink[0]{}%
\providecommand \url  [0]{\begingroup\@sanitize@url \@url }%
\providecommand \@url [1]{\endgroup\@href {#1}{\urlprefix }}%
\providecommand \urlprefix  [0]{URL }%
\providecommand \Eprint [0]{\href }%
\providecommand \doibase [0]{https://doi.org/}%
\providecommand \selectlanguage [0]{\@gobble}%
\providecommand \bibinfo  [0]{\@secondoftwo}%
\providecommand \bibfield  [0]{\@secondoftwo}%
\providecommand \translation [1]{[#1]}%
\providecommand \BibitemOpen [0]{}%
\providecommand \bibitemStop [0]{}%
\providecommand \bibitemNoStop [0]{.\EOS\space}%
\providecommand \EOS [0]{\spacefactor3000\relax}%
\providecommand \BibitemShut  [1]{\csname bibitem#1\endcsname}%
\let\auto@bib@innerbib\@empty
\bibitem [{\citenamefont {De~Nicola}\ \emph {et~al.}(2021)\citenamefont
  {De~Nicola}, \citenamefont {Michailidis},\ and\ \citenamefont
  {Serbyn}}]{entanglementView}%
  \BibitemOpen
  \bibfield  {author} {\bibinfo {author} {\bibfnamefont {S.}~\bibnamefont
  {De~Nicola}}, \bibinfo {author} {\bibfnamefont {A.~A.}\ \bibnamefont
  {Michailidis}},\ and\ \bibinfo {author} {\bibfnamefont {M.}~\bibnamefont
  {Serbyn}},\ }\bibfield  {title} {\bibinfo {title} {Entanglement view of
  dynamical quantum phase transitions},\ }\href
  {https://doi.org/10.1103/PhysRevLett.126.040602} {\bibfield  {journal}
  {\bibinfo  {journal} {Phys. Rev. Lett.}\ }\textbf {\bibinfo {volume} {126}},\
  \bibinfo {pages} {040602} (\bibinfo {year} {2021})}\BibitemShut {NoStop}%
\bibitem [{\citenamefont {Georgescu}\ \emph {et~al.}(2014)\citenamefont
  {Georgescu}, \citenamefont {Ashhab},\ and\ \citenamefont
  {Nori}}]{Georgescu2014}%
  \BibitemOpen
  \bibfield  {author} {\bibinfo {author} {\bibfnamefont {I.~M.}\ \bibnamefont
  {Georgescu}}, \bibinfo {author} {\bibfnamefont {S.}~\bibnamefont {Ashhab}},\
  and\ \bibinfo {author} {\bibfnamefont {F.}~\bibnamefont {Nori}},\ }\bibfield
  {title} {\bibinfo {title} {Quantum simulation},\ }\href
  {https://doi.org/10.1103/RevModPhys.86.153} {\bibfield  {journal} {\bibinfo
  {journal} {Rev. Mod. Phys.}\ }\textbf {\bibinfo {volume} {86}},\ \bibinfo
  {pages} {153} (\bibinfo {year} {2014})}\BibitemShut {NoStop}%
\bibitem [{\citenamefont {Blatt}\ and\ \citenamefont
  {Roos}(2012)}]{blattRoos2012}%
  \BibitemOpen
  \bibfield  {author} {\bibinfo {author} {\bibfnamefont {R.}~\bibnamefont
  {Blatt}}\ and\ \bibinfo {author} {\bibfnamefont {C.~F.}\ \bibnamefont
  {Roos}},\ }\bibfield  {title} {\bibinfo {title} {Quantum simulations with
  trapped ions},\ }\href {http://dx.doi.org/10.1038/nphys2252} {\bibfield
  {journal} {\bibinfo  {journal} {Nature Phys.}\ }\textbf {\bibinfo {volume}
  {8}},\ \bibinfo {pages} {277} (\bibinfo {year} {2012})}\BibitemShut {NoStop}%
\bibitem [{\citenamefont {Schneider}\ \emph {et~al.}(2012)\citenamefont
  {Schneider}, \citenamefont {Porras},\ and\ \citenamefont
  {Schaetz}}]{Schneider_2012}%
  \BibitemOpen
  \bibfield  {author} {\bibinfo {author} {\bibfnamefont {C.}~\bibnamefont
  {Schneider}}, \bibinfo {author} {\bibfnamefont {D.}~\bibnamefont {Porras}},\
  and\ \bibinfo {author} {\bibfnamefont {T.}~\bibnamefont {Schaetz}},\
  }\bibfield  {title} {\bibinfo {title} {Experimental quantum simulations of
  many-body physics with trapped ions},\ }\href
  {https://doi.org/10.1088/0034-4885/75/2/024401} {\bibfield  {journal}
  {\bibinfo  {journal} {Rep. Prog. Phys.}\ }\textbf {\bibinfo {volume} {75}},\
  \bibinfo {pages} {024401} (\bibinfo {year} {2012})}\BibitemShut {NoStop}%
\bibitem [{\citenamefont {Langen}\ \emph {et~al.}(2015)\citenamefont {Langen},
  \citenamefont {Geiger},\ and\ \citenamefont
  {Schmiedmayer}}]{reviewColdAtoms2015}%
  \BibitemOpen
  \bibfield  {author} {\bibinfo {author} {\bibfnamefont {T.}~\bibnamefont
  {Langen}}, \bibinfo {author} {\bibfnamefont {R.}~\bibnamefont {Geiger}},\
  and\ \bibinfo {author} {\bibfnamefont {J.}~\bibnamefont {Schmiedmayer}},\
  }\bibfield  {title} {\bibinfo {title} {Ultracold atoms out of equilibrium},\
  }\href {https://doi.org/10.1146/annurev-conmatphys-031214-014548} {\bibfield
  {journal} {\bibinfo  {journal} {Annu. Rev. Condens. Matter Phys.}\ }\textbf
  {\bibinfo {volume} {6}},\ \bibinfo {pages} {201} (\bibinfo {year}
  {2015})}\BibitemShut {NoStop}%
\bibitem [{\citenamefont {Gross}\ and\ \citenamefont
  {Bloch}(2017)}]{Gross2017}%
  \BibitemOpen
  \bibfield  {author} {\bibinfo {author} {\bibfnamefont {C.}~\bibnamefont
  {Gross}}\ and\ \bibinfo {author} {\bibfnamefont {I.}~\bibnamefont {Bloch}},\
  }\bibfield  {title} {\bibinfo {title} {Quantum simulations with ultracold
  atoms in optical lattices},\ }\href {https://doi.org/10.1126/science.aal3837}
  {\bibfield  {journal} {\bibinfo  {journal} {Science}\ }\textbf {\bibinfo
  {volume} {357}},\ \bibinfo {pages} {995} (\bibinfo {year}
  {2017})}\BibitemShut {NoStop}%
\bibitem [{\citenamefont {Eisert}\ \emph {et~al.}(2015)\citenamefont {Eisert},
  \citenamefont {Friesdorf},\ and\ \citenamefont {Gogolin}}]{eisert2015}%
  \BibitemOpen
  \bibfield  {author} {\bibinfo {author} {\bibfnamefont {J.}~\bibnamefont
  {Eisert}}, \bibinfo {author} {\bibfnamefont {M.}~\bibnamefont {Friesdorf}},\
  and\ \bibinfo {author} {\bibfnamefont {C.}~\bibnamefont {Gogolin}},\
  }\bibfield  {title} {\bibinfo {title} {Quantum many-body systems out of
  equilibrium},\ }\href {http://dx.doi.org/10.1038/nphys3215} {\bibfield
  {journal} {\bibinfo  {journal} {Nature Phys.}\ }\textbf {\bibinfo {volume}
  {11}},\ \bibinfo {pages} {124} (\bibinfo {year} {2015})}\BibitemShut
  {NoStop}%
\bibitem [{\citenamefont {Nandkishore}\ and\ \citenamefont
  {Huse}(2015)}]{Nandkishore2015}%
  \BibitemOpen
  \bibfield  {author} {\bibinfo {author} {\bibfnamefont {R.}~\bibnamefont
  {Nandkishore}}\ and\ \bibinfo {author} {\bibfnamefont {D.~A.}\ \bibnamefont
  {Huse}},\ }\bibfield  {title} {\bibinfo {title} {Many-body localization and
  thermalization in quantum statistical mechanics},\ }\href
  {https://doi.org/10.1146/annurev-conmatphys-031214-014726} {\bibfield
  {journal} {\bibinfo  {journal} {Annu. Rev. Condens. Matter Phys.}\ }\textbf
  {\bibinfo {volume} {6}},\ \bibinfo {pages} {15} (\bibinfo {year} {2015})},\
  \Eprint
  {https://arxiv.org/abs/https://doi.org/10.1146/annurev-conmatphys-031214-014726}
  {https://doi.org/10.1146/annurev-conmatphys-031214-014726} \BibitemShut
  {NoStop}%
\bibitem [{\citenamefont {Abanin}\ \emph {et~al.}(2019)\citenamefont {Abanin},
  \citenamefont {Altman}, \citenamefont {Bloch},\ and\ \citenamefont
  {Serbyn}}]{Abanin2019}%
  \BibitemOpen
  \bibfield  {author} {\bibinfo {author} {\bibfnamefont {D.~A.}\ \bibnamefont
  {Abanin}}, \bibinfo {author} {\bibfnamefont {E.}~\bibnamefont {Altman}},
  \bibinfo {author} {\bibfnamefont {I.}~\bibnamefont {Bloch}},\ and\ \bibinfo
  {author} {\bibfnamefont {M.}~\bibnamefont {Serbyn}},\ }\bibfield  {title}
  {\bibinfo {title} {Colloquium: Many-body localization, thermalization, and
  entanglement},\ }\href {https://doi.org/10.1103/RevModPhys.91.021001}
  {\bibfield  {journal} {\bibinfo  {journal} {Rev. Mod. Phys.}\ }\textbf
  {\bibinfo {volume} {91}},\ \bibinfo {pages} {021001} (\bibinfo {year}
  {2019})}\BibitemShut {NoStop}%
\bibitem [{\citenamefont {Turner}\ \emph {et~al.}(2018)\citenamefont {Turner},
  \citenamefont {Michailidis}, \citenamefont {Abanin}, \citenamefont {Serbyn},\
  and\ \citenamefont {Papi{\'{c}}}}]{Turner2018}%
  \BibitemOpen
  \bibfield  {author} {\bibinfo {author} {\bibfnamefont {C.~J.}\ \bibnamefont
  {Turner}}, \bibinfo {author} {\bibfnamefont {A.~A.}\ \bibnamefont
  {Michailidis}}, \bibinfo {author} {\bibfnamefont {D.~A.}\ \bibnamefont
  {Abanin}}, \bibinfo {author} {\bibfnamefont {M.}~\bibnamefont {Serbyn}},\
  and\ \bibinfo {author} {\bibfnamefont {Z.}~\bibnamefont {Papi{\'{c}}}},\
  }\bibfield  {title} {\bibinfo {title} {Weak ergodicity breaking from quantum
  many-body scars},\ }\href {https://doi.org/10.1038/s41567-018-0137-5}
  {\bibfield  {journal} {\bibinfo  {journal} {Nature Physics}\ }\textbf
  {\bibinfo {volume} {14}},\ \bibinfo {pages} {745} (\bibinfo {year}
  {2018})}\BibitemShut {NoStop}%
\bibitem [{\citenamefont {Serbyn}\ \emph {et~al.}(2021)\citenamefont {Serbyn},
  \citenamefont {Abanin},\ and\ \citenamefont {Papi{\'{c}}}}]{Serbyn2021}%
  \BibitemOpen
  \bibfield  {author} {\bibinfo {author} {\bibfnamefont {M.}~\bibnamefont
  {Serbyn}}, \bibinfo {author} {\bibfnamefont {D.~A.}\ \bibnamefont {Abanin}},\
  and\ \bibinfo {author} {\bibfnamefont {Z.}~\bibnamefont {Papi{\'{c}}}},\
  }\bibfield  {title} {\bibinfo {title} {Quantum many-body scars and weak
  breaking of ergodicity},\ }\href {https://doi.org/10.1038/s41567-021-01230-2}
  {\bibfield  {journal} {\bibinfo  {journal} {Nature Physics}\ }\textbf
  {\bibinfo {volume} {17}},\ \bibinfo {pages} {675} (\bibinfo {year}
  {2021})}\BibitemShut {NoStop}%
\bibitem [{\citenamefont {Castro-Alvaredo}\ \emph {et~al.}(2016)\citenamefont
  {Castro-Alvaredo}, \citenamefont {Doyon},\ and\ \citenamefont
  {Yoshimura}}]{doyon2016}%
  \BibitemOpen
  \bibfield  {author} {\bibinfo {author} {\bibfnamefont {O.~A.}\ \bibnamefont
  {Castro-Alvaredo}}, \bibinfo {author} {\bibfnamefont {B.}~\bibnamefont
  {Doyon}},\ and\ \bibinfo {author} {\bibfnamefont {T.}~\bibnamefont
  {Yoshimura}},\ }\bibfield  {title} {\bibinfo {title} {Emergent hydrodynamics
  in integrable quantum systems out of equilibrium},\ }\href
  {https://doi.org/10.1103/PhysRevX.6.041065} {\bibfield  {journal} {\bibinfo
  {journal} {Phys. Rev. X}\ }\textbf {\bibinfo {volume} {6}},\ \bibinfo {pages}
  {041065} (\bibinfo {year} {2016})}\BibitemShut {NoStop}%
\bibitem [{\citenamefont {Bertini}\ \emph {et~al.}(2016)\citenamefont
  {Bertini}, \citenamefont {Collura}, \citenamefont {De~Nardis},\ and\
  \citenamefont {Fagotti}}]{bertini2016}%
  \BibitemOpen
  \bibfield  {author} {\bibinfo {author} {\bibfnamefont {B.}~\bibnamefont
  {Bertini}}, \bibinfo {author} {\bibfnamefont {M.}~\bibnamefont {Collura}},
  \bibinfo {author} {\bibfnamefont {J.}~\bibnamefont {De~Nardis}},\ and\
  \bibinfo {author} {\bibfnamefont {M.}~\bibnamefont {Fagotti}},\ }\bibfield
  {title} {\bibinfo {title} {Transport in out-of-equilibrium {XXZ} chains:
  Exact profiles of charges and currents},\ }\href
  {https://doi.org/10.1103/PhysRevLett.117.207201} {\bibfield  {journal}
  {\bibinfo  {journal} {Phys. Rev. Lett.}\ }\textbf {\bibinfo {volume} {117}},\
  \bibinfo {pages} {207201} (\bibinfo {year} {2016})}\BibitemShut {NoStop}%
\bibitem [{\citenamefont {Else}\ \emph {et~al.}(2020)\citenamefont {Else},
  \citenamefont {Monroe}, \citenamefont {Nayak},\ and\ \citenamefont
  {Yao}}]{Else2020}%
  \BibitemOpen
  \bibfield  {author} {\bibinfo {author} {\bibfnamefont {D.~V.}\ \bibnamefont
  {Else}}, \bibinfo {author} {\bibfnamefont {C.}~\bibnamefont {Monroe}},
  \bibinfo {author} {\bibfnamefont {C.}~\bibnamefont {Nayak}},\ and\ \bibinfo
  {author} {\bibfnamefont {N.~Y.}\ \bibnamefont {Yao}},\ }\bibfield  {title}
  {\bibinfo {title} {Discrete time crystals},\ }\href
  {https://doi.org/10.1146/annurev-conmatphys-031119-050658} {\bibfield
  {journal} {\bibinfo  {journal} {Annu. Rev. Condens. Matter Phys.}\ }\textbf
  {\bibinfo {volume} {11}},\ \bibinfo {pages} {467} (\bibinfo {year} {2020})},\
  \Eprint
  {https://arxiv.org/abs/https://doi.org/10.1146/annurev-conmatphys-031119-050658}
  {https://doi.org/10.1146/annurev-conmatphys-031119-050658} \BibitemShut
  {NoStop}%
\bibitem [{\citenamefont {Calabrese}\ and\ \citenamefont
  {Cardy}(2005)}]{Calabrese_2005}%
  \BibitemOpen
  \bibfield  {author} {\bibinfo {author} {\bibfnamefont {P.}~\bibnamefont
  {Calabrese}}\ and\ \bibinfo {author} {\bibfnamefont {J.}~\bibnamefont
  {Cardy}},\ }\bibfield  {title} {\bibinfo {title} {Evolution of entanglement
  entropy in one-dimensional systems},\ }\href
  {https://doi.org/10.1088/1742-5468/2005/04/p04010} {\bibfield  {journal}
  {\bibinfo  {journal} {J. Stat. Mech. Theor. Exp.}\ }\textbf {\bibinfo
  {volume} {2005}},\ \bibinfo {pages} {P04010} (\bibinfo {year}
  {2005})}\BibitemShut {NoStop}%
\bibitem [{\citenamefont {Calabrese}\ and\ \citenamefont
  {Cardy}(2006)}]{Calabrese2006}%
  \BibitemOpen
  \bibfield  {author} {\bibinfo {author} {\bibfnamefont {P.}~\bibnamefont
  {Calabrese}}\ and\ \bibinfo {author} {\bibfnamefont {J.}~\bibnamefont
  {Cardy}},\ }\bibfield  {title} {\bibinfo {title} {Time dependence of
  correlation functions following a quantum quench},\ }\href
  {https://doi.org/10.1103/PhysRevLett.96.136801} {\bibfield  {journal}
  {\bibinfo  {journal} {Phys. Rev. Lett.}\ }\textbf {\bibinfo {volume} {96}},\
  \bibinfo {pages} {136801} (\bibinfo {year} {2006})}\BibitemShut {NoStop}%
\bibitem [{\citenamefont {Essler}\ and\ \citenamefont
  {Fagotti}(2016)}]{Essler_2016}%
  \BibitemOpen
  \bibfield  {author} {\bibinfo {author} {\bibfnamefont {F.~H.~L.}\
  \bibnamefont {Essler}}\ and\ \bibinfo {author} {\bibfnamefont
  {M.}~\bibnamefont {Fagotti}},\ }\bibfield  {title} {\bibinfo {title} {Quench
  dynamics and relaxation in isolated integrable quantum spin chains},\ }\href
  {https://doi.org/10.1088/1742-5468/2016/06/064002} {\bibfield  {journal}
  {\bibinfo  {journal} {J. Stat. Mech.: Theory Exp.}\ }\textbf {\bibinfo
  {volume} {2016}}\bibinfo  {number} { (6)},\ \bibinfo {pages}
  {064002}}\BibitemShut {NoStop}%
\bibitem [{\citenamefont {Mitra}(2018)}]{Mitra2018}%
  \BibitemOpen
\bibfield  {number} {  }\bibfield  {author} {\bibinfo {author} {\bibfnamefont
  {A.}~\bibnamefont {Mitra}},\ }\bibfield  {title} {\bibinfo {title} {Quantum
  quench dynamics},\ }\href
  {https://doi.org/10.1146/annurev-conmatphys-031016-025451} {\bibfield
  {journal} {\bibinfo  {journal} {Annu. Rev. Condens. Matter Phys.}\ }\textbf
  {\bibinfo {volume} {9}},\ \bibinfo {pages} {245} (\bibinfo {year}
  {2018})}\BibitemShut {NoStop}%
\bibitem [{\citenamefont {Paeckel}\ \emph {et~al.}(2019)\citenamefont
  {Paeckel}, \citenamefont {K{\"{o}}hler}, \citenamefont {Swoboda},
  \citenamefont {Manmana}, \citenamefont {Schollw{\"{o}}ck},\ and\
  \citenamefont {Hubig}}]{Paeckel2019}%
  \BibitemOpen
  \bibfield  {author} {\bibinfo {author} {\bibfnamefont {S.}~\bibnamefont
  {Paeckel}}, \bibinfo {author} {\bibfnamefont {T.}~\bibnamefont
  {K{\"{o}}hler}}, \bibinfo {author} {\bibfnamefont {A.}~\bibnamefont
  {Swoboda}}, \bibinfo {author} {\bibfnamefont {S.~R.}\ \bibnamefont
  {Manmana}}, \bibinfo {author} {\bibfnamefont {U.}~\bibnamefont
  {Schollw{\"{o}}ck}},\ and\ \bibinfo {author} {\bibfnamefont {C.}~\bibnamefont
  {Hubig}},\ }\bibfield  {title} {\bibinfo {title} {{Time-evolution methods for
  matrix-product states}},\ }\href
  {https://doi.org/https://doi.org/10.1016/j.aop.2019.167998} {\bibfield
  {journal} {\bibinfo  {journal} {Ann. Phys. (N. Y).}\ }\textbf {\bibinfo
  {volume} {411}},\ \bibinfo {pages} {167998} (\bibinfo {year}
  {2019})}\BibitemShut {NoStop}%
\bibitem [{\citenamefont {Heyl}\ \emph {et~al.}(2013)\citenamefont {Heyl},
  \citenamefont {Polkovnikov},\ and\ \citenamefont {Kehrein}}]{heyl2013}%
  \BibitemOpen
  \bibfield  {author} {\bibinfo {author} {\bibfnamefont {M.}~\bibnamefont
  {Heyl}}, \bibinfo {author} {\bibfnamefont {A.}~\bibnamefont {Polkovnikov}},\
  and\ \bibinfo {author} {\bibfnamefont {S.}~\bibnamefont {Kehrein}},\
  }\bibfield  {title} {\bibinfo {title} {{Dynamical quantum phase transitions
  in the transverse-field Ising model}},\ }\href
  {https://doi.org/10.1103/PhysRevLett.110.135704} {\bibfield  {journal}
  {\bibinfo  {journal} {Phys. Rev. Lett.}\ }\textbf {\bibinfo {volume} {110}},\
  \bibinfo {pages} {135704} (\bibinfo {year} {2013})}\BibitemShut {NoStop}%
\bibitem [{\citenamefont {Heyl}(2018)}]{heyl2018}%
  \BibitemOpen
  \bibfield  {author} {\bibinfo {author} {\bibfnamefont {M.}~\bibnamefont
  {Heyl}},\ }\bibfield  {title} {\bibinfo {title} {Dynamical quantum phase
  transitions: a review},\ }\href {https://doi.org/10.1088/1361-6633/aaaf9a}
  {\bibfield  {journal} {\bibinfo  {journal} {Rep. Prog. Phys.}\ }\textbf
  {\bibinfo {volume} {81}},\ \bibinfo {pages} {054001} (\bibinfo {year}
  {2018})}\BibitemShut {NoStop}%
\bibitem [{\citenamefont {Karrasch}\ and\ \citenamefont
  {Schuricht}(2013)}]{KarraschSchuricht2013}%
  \BibitemOpen
  \bibfield  {author} {\bibinfo {author} {\bibfnamefont {C.}~\bibnamefont
  {Karrasch}}\ and\ \bibinfo {author} {\bibfnamefont {D.}~\bibnamefont
  {Schuricht}},\ }\bibfield  {title} {\bibinfo {title} {{Dynamical phase
  transitions after quenches in nonintegrable models}},\ }\href
  {https://doi.org/10.1103/PhysRevB.87.195104} {\bibfield  {journal} {\bibinfo
  {journal} {Phys. Rev. B}\ }\textbf {\bibinfo {volume} {87}},\ \bibinfo
  {pages} {195104} (\bibinfo {year} {2013})}\BibitemShut {NoStop}%
\bibitem [{\citenamefont {Canovi}\ \emph
  {et~al.}(2014{\natexlab{a}})\citenamefont {Canovi}, \citenamefont {Werner},\
  and\ \citenamefont {Eckstein}}]{canovi2014}%
  \BibitemOpen
  \bibfield  {author} {\bibinfo {author} {\bibfnamefont {E.}~\bibnamefont
  {Canovi}}, \bibinfo {author} {\bibfnamefont {P.}~\bibnamefont {Werner}},\
  and\ \bibinfo {author} {\bibfnamefont {M.}~\bibnamefont {Eckstein}},\
  }\bibfield  {title} {\bibinfo {title} {First-order dynamical phase
  transitions},\ }\href {https://doi.org/10.1103/PhysRevLett.113.265702}
  {\bibfield  {journal} {\bibinfo  {journal} {Phys. Rev. Lett.}\ }\textbf
  {\bibinfo {volume} {113}},\ \bibinfo {pages} {265702} (\bibinfo {year}
  {2014}{\natexlab{a}})}\BibitemShut {NoStop}%
\bibitem [{\citenamefont {Vajna}\ and\ \citenamefont
  {D\'ora}(2015)}]{vajnaDora2015}%
  \BibitemOpen
  \bibfield  {author} {\bibinfo {author} {\bibfnamefont {S.}~\bibnamefont
  {Vajna}}\ and\ \bibinfo {author} {\bibfnamefont {B.}~\bibnamefont {D\'ora}},\
  }\bibfield  {title} {\bibinfo {title} {Topological classification of
  dynamical phase transitions},\ }\href
  {https://doi.org/10.1103/PhysRevB.91.155127} {\bibfield  {journal} {\bibinfo
  {journal} {Phys. Rev. B}\ }\textbf {\bibinfo {volume} {91}},\ \bibinfo
  {pages} {155127} (\bibinfo {year} {2015})}\BibitemShut {NoStop}%
\bibitem [{\citenamefont {Schmitt}\ and\ \citenamefont
  {Kehrein}(2015)}]{schmittKehrein2015}%
  \BibitemOpen
  \bibfield  {author} {\bibinfo {author} {\bibfnamefont {M.}~\bibnamefont
  {Schmitt}}\ and\ \bibinfo {author} {\bibfnamefont {S.}~\bibnamefont
  {Kehrein}},\ }\bibfield  {title} {\bibinfo {title} {{Dynamical quantum phase
  transitions in the Kitaev honeycomb model}},\ }\href
  {https://doi.org/10.1103/PhysRevB.92.075114} {\bibfield  {journal} {\bibinfo
  {journal} {Phys. Rev. B}\ }\textbf {\bibinfo {volume} {92}},\ \bibinfo
  {pages} {075114} (\bibinfo {year} {2015})}\BibitemShut {NoStop}%
\bibitem [{\citenamefont {Homrighausen}\ \emph {et~al.}(2017)\citenamefont
  {Homrighausen}, \citenamefont {Abeling}, \citenamefont {Zauner-Stauber},\
  and\ \citenamefont {Halimeh}}]{Homrighausen2017}%
  \BibitemOpen
  \bibfield  {author} {\bibinfo {author} {\bibfnamefont {I.}~\bibnamefont
  {Homrighausen}}, \bibinfo {author} {\bibfnamefont {N.~O.}\ \bibnamefont
  {Abeling}}, \bibinfo {author} {\bibfnamefont {V.}~\bibnamefont
  {Zauner-Stauber}},\ and\ \bibinfo {author} {\bibfnamefont {J.~C.}\
  \bibnamefont {Halimeh}},\ }\bibfield  {title} {\bibinfo {title} {Anomalous
  dynamical phase in quantum spin chains with long-range interactions},\ }\href
  {https://doi.org/10.1103/PhysRevB.96.104436} {\bibfield  {journal} {\bibinfo
  {journal} {Phys. Rev. B}\ }\textbf {\bibinfo {volume} {96}},\ \bibinfo
  {pages} {104436} (\bibinfo {year} {2017})}\BibitemShut {NoStop}%
\bibitem [{\citenamefont {\ifmmode \check{Z}\else
  \v{Z}\fi{}unkovi\ifmmode~\check{c}\else \v{c}\fi{}}\ \emph
  {et~al.}(2018)\citenamefont {\ifmmode \check{Z}\else
  \v{Z}\fi{}unkovi\ifmmode~\check{c}\else \v{c}\fi{}}, \citenamefont {Heyl},
  \citenamefont {Knap},\ and\ \citenamefont {Silva}}]{zunkovic2018}%
  \BibitemOpen
  \bibfield  {author} {\bibinfo {author} {\bibfnamefont {B.}~\bibnamefont
  {\ifmmode \check{Z}\else \v{Z}\fi{}unkovi\ifmmode~\check{c}\else
  \v{c}\fi{}}}, \bibinfo {author} {\bibfnamefont {M.}~\bibnamefont {Heyl}},
  \bibinfo {author} {\bibfnamefont {M.}~\bibnamefont {Knap}},\ and\ \bibinfo
  {author} {\bibfnamefont {A.}~\bibnamefont {Silva}},\ }\bibfield  {title}
  {\bibinfo {title} {Dynamical quantum phase transitions in spin chains with
  long-range interactions: Merging different concepts of nonequilibrium
  criticality},\ }\href {https://doi.org/10.1103/PhysRevLett.120.130601}
  {\bibfield  {journal} {\bibinfo  {journal} {Phys. Rev. Lett.}\ }\textbf
  {\bibinfo {volume} {120}},\ \bibinfo {pages} {130601} (\bibinfo {year}
  {2018})}\BibitemShut {NoStop}%
\bibitem [{\citenamefont {Gurarie}(2019)}]{Gurarie2019}%
  \BibitemOpen
  \bibfield  {author} {\bibinfo {author} {\bibfnamefont {V.}~\bibnamefont
  {Gurarie}},\ }\bibfield  {title} {\bibinfo {title} {{Dynamical quantum phase
  transitions in the random field Ising model}},\ }\href
  {https://doi.org/10.1103/PhysRevA.100.031601} {\bibfield  {journal} {\bibinfo
   {journal} {Phys. Rev. A}\ }\textbf {\bibinfo {volume} {100}},\ \bibinfo
  {pages} {031601} (\bibinfo {year} {2019})}\BibitemShut {NoStop}%
\bibitem [{\citenamefont {Huang}\ \emph {et~al.}(2019)\citenamefont {Huang},
  \citenamefont {Banerjee},\ and\ \citenamefont {Heyl}}]{Huang2019}%
  \BibitemOpen
  \bibfield  {author} {\bibinfo {author} {\bibfnamefont {Y.-P.}\ \bibnamefont
  {Huang}}, \bibinfo {author} {\bibfnamefont {D.}~\bibnamefont {Banerjee}},\
  and\ \bibinfo {author} {\bibfnamefont {M.}~\bibnamefont {Heyl}},\ }\bibfield
  {title} {\bibinfo {title} {Dynamical quantum phase transitions in {U(1)}
  quantum link models},\ }\href
  {https://doi.org/10.1103/PhysRevLett.122.250401} {\bibfield  {journal}
  {\bibinfo  {journal} {Phys. Rev. Lett.}\ }\textbf {\bibinfo {volume} {122}},\
  \bibinfo {pages} {250401} (\bibinfo {year} {2019})}\BibitemShut {NoStop}%
\bibitem [{\citenamefont {Hagym\'asi}\ \emph {et~al.}(2019)\citenamefont
  {Hagym\'asi}, \citenamefont {Hubig}, \citenamefont {Legeza},\ and\
  \citenamefont {Schollw\"ock}}]{Hagymasi2019}%
  \BibitemOpen
  \bibfield  {author} {\bibinfo {author} {\bibfnamefont {I.}~\bibnamefont
  {Hagym\'asi}}, \bibinfo {author} {\bibfnamefont {C.}~\bibnamefont {Hubig}},
  \bibinfo {author} {\bibfnamefont {O.}~\bibnamefont {Legeza}},\ and\ \bibinfo
  {author} {\bibfnamefont {U.}~\bibnamefont {Schollw\"ock}},\ }\bibfield
  {title} {\bibinfo {title} {Dynamical topological quantum phase transitions in
  nonintegrable models},\ }\href
  {https://doi.org/10.1103/PhysRevLett.122.250601} {\bibfield  {journal}
  {\bibinfo  {journal} {Phys. Rev. Lett.}\ }\textbf {\bibinfo {volume} {122}},\
  \bibinfo {pages} {250601} (\bibinfo {year} {2019})}\BibitemShut {NoStop}%
\bibitem [{\citenamefont {Jurcevic}\ \emph {et~al.}(2017)\citenamefont
  {Jurcevic}, \citenamefont {Shen}, \citenamefont {Hauke}, \citenamefont
  {Maier}, \citenamefont {Brydges}, \citenamefont {Hempel}, \citenamefont
  {Lanyon}, \citenamefont {Heyl}, \citenamefont {Blatt},\ and\ \citenamefont
  {Roos}}]{jurcevic2017}%
  \BibitemOpen
  \bibfield  {author} {\bibinfo {author} {\bibfnamefont {P.}~\bibnamefont
  {Jurcevic}}, \bibinfo {author} {\bibfnamefont {H.}~\bibnamefont {Shen}},
  \bibinfo {author} {\bibfnamefont {P.}~\bibnamefont {Hauke}}, \bibinfo
  {author} {\bibfnamefont {C.}~\bibnamefont {Maier}}, \bibinfo {author}
  {\bibfnamefont {T.}~\bibnamefont {Brydges}}, \bibinfo {author} {\bibfnamefont
  {C.}~\bibnamefont {Hempel}}, \bibinfo {author} {\bibfnamefont {B.~P.}\
  \bibnamefont {Lanyon}}, \bibinfo {author} {\bibfnamefont {M.}~\bibnamefont
  {Heyl}}, \bibinfo {author} {\bibfnamefont {R.}~\bibnamefont {Blatt}},\ and\
  \bibinfo {author} {\bibfnamefont {C.~F.}\ \bibnamefont {Roos}},\ }\bibfield
  {title} {\bibinfo {title} {Direct observation of dynamical quantum phase
  transitions in an interacting many-body system},\ }\href
  {https://doi.org/10.1103/PhysRevLett.119.080501} {\bibfield  {journal}
  {\bibinfo  {journal} {Phys. Rev. Lett.}\ }\textbf {\bibinfo {volume} {119}},\
  \bibinfo {pages} {080501} (\bibinfo {year} {2017})}\BibitemShut {NoStop}%
\bibitem [{\citenamefont {Fl{\"{a}}schner}\ \emph {et~al.}(2018)\citenamefont
  {Fl{\"{a}}schner}, \citenamefont {Vogel}, \citenamefont {Tarnowski},
  \citenamefont {Rem}, \citenamefont {L{\"{u}}hmann}, \citenamefont {Heyl},
  \citenamefont {Budich}, \citenamefont {Mathey}, \citenamefont {Sengstock},\
  and\ \citenamefont {Weitenberg}}]{Flaschner2018}%
  \BibitemOpen
  \bibfield  {author} {\bibinfo {author} {\bibfnamefont {N.}~\bibnamefont
  {Fl{\"{a}}schner}}, \bibinfo {author} {\bibfnamefont {D.}~\bibnamefont
  {Vogel}}, \bibinfo {author} {\bibfnamefont {M.}~\bibnamefont {Tarnowski}},
  \bibinfo {author} {\bibfnamefont {B.~S.}\ \bibnamefont {Rem}}, \bibinfo
  {author} {\bibfnamefont {D.-S.}\ \bibnamefont {L{\"{u}}hmann}}, \bibinfo
  {author} {\bibfnamefont {M.}~\bibnamefont {Heyl}}, \bibinfo {author}
  {\bibfnamefont {J.~C.}\ \bibnamefont {Budich}}, \bibinfo {author}
  {\bibfnamefont {L.}~\bibnamefont {Mathey}}, \bibinfo {author} {\bibfnamefont
  {K.}~\bibnamefont {Sengstock}},\ and\ \bibinfo {author} {\bibfnamefont
  {C.}~\bibnamefont {Weitenberg}},\ }\bibfield  {title} {\bibinfo {title}
  {{Observation of dynamical vortices after quenches in a system with
  topology}},\ }\href {https://doi.org/10.1038/s41567-017-0013-8} {\bibfield
  {journal} {\bibinfo  {journal} {Nat. Phys.}\ }\textbf {\bibinfo {volume}
  {14}},\ \bibinfo {pages} {265} (\bibinfo {year} {2018})}\BibitemShut
  {NoStop}%
\bibitem [{\citenamefont {Tian}\ \emph {et~al.}(2019)\citenamefont {Tian},
  \citenamefont {Ke}, \citenamefont {Zhang}, \citenamefont {Lin}, \citenamefont
  {Shi}, \citenamefont {Huang}, \citenamefont {Lee},\ and\ \citenamefont
  {Du}}]{Tian2018}%
  \BibitemOpen
  \bibfield  {author} {\bibinfo {author} {\bibfnamefont {T.}~\bibnamefont
  {Tian}}, \bibinfo {author} {\bibfnamefont {Y.}~\bibnamefont {Ke}}, \bibinfo
  {author} {\bibfnamefont {L.}~\bibnamefont {Zhang}}, \bibinfo {author}
  {\bibfnamefont {S.}~\bibnamefont {Lin}}, \bibinfo {author} {\bibfnamefont
  {Z.}~\bibnamefont {Shi}}, \bibinfo {author} {\bibfnamefont {P.}~\bibnamefont
  {Huang}}, \bibinfo {author} {\bibfnamefont {C.}~\bibnamefont {Lee}},\ and\
  \bibinfo {author} {\bibfnamefont {J.}~\bibnamefont {Du}},\ }\bibfield
  {title} {\bibinfo {title} {Observation of dynamical phase transitions in a
  topological nanomechanical system},\ }\href
  {https://doi.org/10.1103/PhysRevB.100.024310} {\bibfield  {journal} {\bibinfo
   {journal} {Phys. Rev. B}\ }\textbf {\bibinfo {volume} {100}},\ \bibinfo
  {pages} {024310} (\bibinfo {year} {2019})}\BibitemShut {NoStop}%
\bibitem [{\citenamefont {Guo}\ \emph {et~al.}(2019)\citenamefont {Guo},
  \citenamefont {Yang}, \citenamefont {Zeng}, \citenamefont {Peng},
  \citenamefont {Li}, \citenamefont {Deng}, \citenamefont {Jin}, \citenamefont
  {Chen}, \citenamefont {Zheng},\ and\ \citenamefont {Fan}}]{Guo2019}%
  \BibitemOpen
  \bibfield  {author} {\bibinfo {author} {\bibfnamefont {X.-Y.}\ \bibnamefont
  {Guo}}, \bibinfo {author} {\bibfnamefont {C.}~\bibnamefont {Yang}}, \bibinfo
  {author} {\bibfnamefont {Y.}~\bibnamefont {Zeng}}, \bibinfo {author}
  {\bibfnamefont {Y.}~\bibnamefont {Peng}}, \bibinfo {author} {\bibfnamefont
  {H.-K.}\ \bibnamefont {Li}}, \bibinfo {author} {\bibfnamefont
  {H.}~\bibnamefont {Deng}}, \bibinfo {author} {\bibfnamefont {Y.-R.}\
  \bibnamefont {Jin}}, \bibinfo {author} {\bibfnamefont {S.}~\bibnamefont
  {Chen}}, \bibinfo {author} {\bibfnamefont {D.}~\bibnamefont {Zheng}},\ and\
  \bibinfo {author} {\bibfnamefont {H.}~\bibnamefont {Fan}},\ }\bibfield
  {title} {\bibinfo {title} {Observation of a dynamical quantum phase
  transition by a superconducting qubit simulation},\ }\href
  {https://doi.org/10.1103/PhysRevApplied.11.044080} {\bibfield  {journal}
  {\bibinfo  {journal} {Phys. Rev. Applied}\ }\textbf {\bibinfo {volume}
  {11}},\ \bibinfo {pages} {044080} (\bibinfo {year} {2019})}\BibitemShut
  {NoStop}%
\bibitem [{\citenamefont {Wang}\ \emph {et~al.}(2019)\citenamefont {Wang},
  \citenamefont {Qiu}, \citenamefont {Xiao}, \citenamefont {Zhan},
  \citenamefont {Bian}, \citenamefont {Yi},\ and\ \citenamefont
  {Xue}}]{Wang2019}%
  \BibitemOpen
  \bibfield  {author} {\bibinfo {author} {\bibfnamefont {K.}~\bibnamefont
  {Wang}}, \bibinfo {author} {\bibfnamefont {X.}~\bibnamefont {Qiu}}, \bibinfo
  {author} {\bibfnamefont {L.}~\bibnamefont {Xiao}}, \bibinfo {author}
  {\bibfnamefont {X.}~\bibnamefont {Zhan}}, \bibinfo {author} {\bibfnamefont
  {Z.}~\bibnamefont {Bian}}, \bibinfo {author} {\bibfnamefont {W.}~\bibnamefont
  {Yi}},\ and\ \bibinfo {author} {\bibfnamefont {P.}~\bibnamefont {Xue}},\
  }\bibfield  {title} {\bibinfo {title} {Simulating dynamic quantum phase
  transitions in photonic quantum walks},\ }\href
  {https://doi.org/10.1103/PhysRevLett.122.020501} {\bibfield  {journal}
  {\bibinfo  {journal} {Phys. Rev. Lett.}\ }\textbf {\bibinfo {volume} {122}},\
  \bibinfo {pages} {020501} (\bibinfo {year} {2019})}\BibitemShut {NoStop}%
\bibitem [{\citenamefont {Torlai}\ \emph {et~al.}(2014)\citenamefont {Torlai},
  \citenamefont {Tagliacozzo},\ and\ \citenamefont {Chiara}}]{Torlai2014}%
  \BibitemOpen
  \bibfield  {author} {\bibinfo {author} {\bibfnamefont {G.}~\bibnamefont
  {Torlai}}, \bibinfo {author} {\bibfnamefont {L.}~\bibnamefont
  {Tagliacozzo}},\ and\ \bibinfo {author} {\bibfnamefont {G.~D.}\ \bibnamefont
  {Chiara}},\ }\bibfield  {title} {\bibinfo {title} {Dynamics of the
  entanglement spectrum in spin chains},\ }\href
  {https://doi.org/10.1088/1742-5468/2014/06/p06001} {\bibfield  {journal}
  {\bibinfo  {journal} {J. Stat. Mech.: Theory Exp.}\ }\textbf {\bibinfo
  {volume} {2014}}\bibinfo  {number} { (6)},\ \bibinfo {pages}
  {P06001}}\BibitemShut {NoStop}%
\bibitem [{\citenamefont {Karrasch}\ and\ \citenamefont
  {Schuricht}(2017)}]{Karrasch2017}%
  \BibitemOpen
\bibfield  {number} {  }\bibfield  {author} {\bibinfo {author} {\bibfnamefont
  {C.}~\bibnamefont {Karrasch}}\ and\ \bibinfo {author} {\bibfnamefont
  {D.}~\bibnamefont {Schuricht}},\ }\bibfield  {title} {\bibinfo {title}
  {{Dynamical quantum phase transitions in the quantum Potts chain}},\ }\href
  {https://doi.org/10.1103/PhysRevB.95.075143} {\bibfield  {journal} {\bibinfo
  {journal} {Phys. Rev. B}\ }\textbf {\bibinfo {volume} {95}},\ \bibinfo
  {pages} {075143} (\bibinfo {year} {2017})}\BibitemShut {NoStop}%
\bibitem [{\citenamefont {Andraschko}\ and\ \citenamefont
  {Sirker}(2014)}]{andraschkoSirker2014}%
  \BibitemOpen
  \bibfield  {author} {\bibinfo {author} {\bibfnamefont {F.}~\bibnamefont
  {Andraschko}}\ and\ \bibinfo {author} {\bibfnamefont {J.}~\bibnamefont
  {Sirker}},\ }\bibfield  {title} {\bibinfo {title} {{Dynamical quantum phase
  transitions and the Loschmidt echo: A transfer matrix approach}},\ }\href
  {https://doi.org/10.1103/PhysRevB.89.125120} {\bibfield  {journal} {\bibinfo
  {journal} {Phys. Rev. B}\ }\textbf {\bibinfo {volume} {89}},\ \bibinfo
  {pages} {125120} (\bibinfo {year} {2014})}\BibitemShut {NoStop}%
\bibitem [{\citenamefont {Vajna}\ and\ \citenamefont
  {D\'ora}(2014)}]{vajnaDora2014}%
  \BibitemOpen
  \bibfield  {author} {\bibinfo {author} {\bibfnamefont {S.}~\bibnamefont
  {Vajna}}\ and\ \bibinfo {author} {\bibfnamefont {B.}~\bibnamefont {D\'ora}},\
  }\bibfield  {title} {\bibinfo {title} {Disentangling dynamical phase
  transitions from equilibrium phase transitions},\ }\href
  {https://doi.org/10.1103/PhysRevB.89.161105} {\bibfield  {journal} {\bibinfo
  {journal} {Phys. Rev. B}\ }\textbf {\bibinfo {volume} {89}},\ \bibinfo
  {pages} {161105} (\bibinfo {year} {2014})}\BibitemShut {NoStop}%
\bibitem [{\citenamefont {Sharma}\ \emph {et~al.}(2015)\citenamefont {Sharma},
  \citenamefont {Suzuki},\ and\ \citenamefont {Dutta}}]{Sharma2015}%
  \BibitemOpen
  \bibfield  {author} {\bibinfo {author} {\bibfnamefont {S.}~\bibnamefont
  {Sharma}}, \bibinfo {author} {\bibfnamefont {S.}~\bibnamefont {Suzuki}},\
  and\ \bibinfo {author} {\bibfnamefont {A.}~\bibnamefont {Dutta}},\ }\bibfield
   {title} {\bibinfo {title} {{Quenches and dynamical phase transitions in a
  nonintegrable quantum Ising model}},\ }\href
  {https://doi.org/10.1103/PhysRevB.92.104306} {\bibfield  {journal} {\bibinfo
  {journal} {Phys. Rev. B}\ }\textbf {\bibinfo {volume} {92}},\ \bibinfo
  {pages} {104306} (\bibinfo {year} {2015})}\BibitemShut {NoStop}%
\bibitem [{\citenamefont {Jafari}(2019)}]{jafari2019}%
  \BibitemOpen
  \bibfield  {author} {\bibinfo {author} {\bibfnamefont {R.}~\bibnamefont
  {Jafari}},\ }\bibfield  {title} {\bibinfo {title} {Dynamical quantum phase
  transition and quasi particle excitation},\ }\href
  {https://doi.org/10.1038/s41598-019-39595-3} {\bibfield  {journal} {\bibinfo
  {journal} {Sci. Rep.}\ }\textbf {\bibinfo {volume} {9}},\ \bibinfo {pages}
  {2871} (\bibinfo {year} {2019})}\BibitemShut {NoStop}%
\bibitem [{\citenamefont {Heyl}(2019)}]{heyl2019}%
  \BibitemOpen
  \bibfield  {author} {\bibinfo {author} {\bibfnamefont {M.}~\bibnamefont
  {Heyl}},\ }\bibfield  {title} {\bibinfo {title} {Dynamical quantum phase
  transitions: A brief survey},\ }\href
  {https://doi.org/10.1209/0295-5075/125/26001} {\bibfield  {journal} {\bibinfo
   {journal} {{EPL}}\ }\textbf {\bibinfo {volume} {125}},\ \bibinfo {pages}
  {26001} (\bibinfo {year} {2019})}\BibitemShut {NoStop}%
\bibitem [{\citenamefont {Heyl}(2014)}]{heyl2014}%
  \BibitemOpen
  \bibfield  {author} {\bibinfo {author} {\bibfnamefont {M.}~\bibnamefont
  {Heyl}},\ }\bibfield  {title} {\bibinfo {title} {Dynamical quantum phase
  transitions in systems with broken-symmetry phases},\ }\href
  {https://doi.org/10.1103/PhysRevLett.113.205701} {\bibfield  {journal}
  {\bibinfo  {journal} {Phys. Rev. Lett.}\ }\textbf {\bibinfo {volume} {113}},\
  \bibinfo {pages} {205701} (\bibinfo {year} {2014})}\BibitemShut {NoStop}%
\bibitem [{\citenamefont {Weidinger}\ \emph {et~al.}(2017)\citenamefont
  {Weidinger}, \citenamefont {Heyl}, \citenamefont {Silva},\ and\ \citenamefont
  {Knap}}]{weidinger2017}%
  \BibitemOpen
  \bibfield  {author} {\bibinfo {author} {\bibfnamefont {S.~A.}\ \bibnamefont
  {Weidinger}}, \bibinfo {author} {\bibfnamefont {M.}~\bibnamefont {Heyl}},
  \bibinfo {author} {\bibfnamefont {A.}~\bibnamefont {Silva}},\ and\ \bibinfo
  {author} {\bibfnamefont {M.}~\bibnamefont {Knap}},\ }\bibfield  {title}
  {\bibinfo {title} {Dynamical quantum phase transitions in systems with
  continuous symmetry breaking},\ }\href
  {https://doi.org/10.1103/PhysRevB.96.134313} {\bibfield  {journal} {\bibinfo
  {journal} {Phys. Rev. B}\ }\textbf {\bibinfo {volume} {96}},\ \bibinfo
  {pages} {134313} (\bibinfo {year} {2017})}\BibitemShut {NoStop}%
\bibitem [{\citenamefont {Feldmeier}\ \emph {et~al.}(2019)\citenamefont
  {Feldmeier}, \citenamefont {Pollmann},\ and\ \citenamefont
  {Knap}}]{feldmeierPollmannKnap2019}%
  \BibitemOpen
  \bibfield  {author} {\bibinfo {author} {\bibfnamefont {J.}~\bibnamefont
  {Feldmeier}}, \bibinfo {author} {\bibfnamefont {F.}~\bibnamefont
  {Pollmann}},\ and\ \bibinfo {author} {\bibfnamefont {M.}~\bibnamefont
  {Knap}},\ }\bibfield  {title} {\bibinfo {title} {Emergent glassy dynamics in
  a quantum dimer model},\ }\href
  {https://doi.org/10.1103/PhysRevLett.123.040601} {\bibfield  {journal}
  {\bibinfo  {journal} {Phys. Rev. Lett.}\ }\textbf {\bibinfo {volume} {123}},\
  \bibinfo {pages} {040601} (\bibinfo {year} {2019})}\BibitemShut {NoStop}%
\bibitem [{\citenamefont {Fogarty}\ \emph {et~al.}(2017)\citenamefont
  {Fogarty}, \citenamefont {Usui}, \citenamefont {Busch}, \citenamefont
  {Silva},\ and\ \citenamefont {Goold}}]{Fogarty_2017}%
  \BibitemOpen
  \bibfield  {author} {\bibinfo {author} {\bibfnamefont {T.}~\bibnamefont
  {Fogarty}}, \bibinfo {author} {\bibfnamefont {A.}~\bibnamefont {Usui}},
  \bibinfo {author} {\bibfnamefont {T.}~\bibnamefont {Busch}}, \bibinfo
  {author} {\bibfnamefont {A.}~\bibnamefont {Silva}},\ and\ \bibinfo {author}
  {\bibfnamefont {J.}~\bibnamefont {Goold}},\ }\bibfield  {title} {\bibinfo
  {title} {Dynamical phase transitions and temporal orthogonality in
  one-dimensional hard-core bosons: from the continuum to the lattice},\ }\href
  {https://doi.org/10.1088/1367-2630/aa8aff} {\bibfield  {journal} {\bibinfo
  {journal} {New J. Phys.}\ }\textbf {\bibinfo {volume} {19}},\ \bibinfo
  {pages} {113018} (\bibinfo {year} {2017})}\BibitemShut {NoStop}%
\bibitem [{\citenamefont {Yu}\ \emph {et~al.}(2021)\citenamefont {Yu},
  \citenamefont {Sacramento}, \citenamefont {Li},\ and\ \citenamefont
  {Lin}}]{yu2021correlations}%
  \BibitemOpen
  \bibfield  {author} {\bibinfo {author} {\bibfnamefont {W.~C.}\ \bibnamefont
  {Yu}}, \bibinfo {author} {\bibfnamefont {P.~D.}\ \bibnamefont {Sacramento}},
  \bibinfo {author} {\bibfnamefont {Y.~C.}\ \bibnamefont {Li}},\ and\ \bibinfo
  {author} {\bibfnamefont {H.-Q.}\ \bibnamefont {Lin}},\ }\href@noop {}
  {\bibinfo {title} {Correlations and dynamical quantum phase transitions in an
  interacting topological insulator}} (\bibinfo {year} {2021}),\ \Eprint
  {https://arxiv.org/abs/2105.02449} {arXiv:2105.02449 [cond-mat.str-el]}
  \BibitemShut {NoStop}%
\bibitem [{\citenamefont {Halimeh}\ \emph {et~al.}(2021)\citenamefont
  {Halimeh}, \citenamefont {Trapin}, \citenamefont {Van~Damme},\ and\
  \citenamefont {Heyl}}]{halimeh2020local}%
  \BibitemOpen
  \bibfield  {author} {\bibinfo {author} {\bibfnamefont {J.~C.}\ \bibnamefont
  {Halimeh}}, \bibinfo {author} {\bibfnamefont {D.}~\bibnamefont {Trapin}},
  \bibinfo {author} {\bibfnamefont {M.}~\bibnamefont {Van~Damme}},\ and\
  \bibinfo {author} {\bibfnamefont {M.}~\bibnamefont {Heyl}},\ }\bibfield
  {title} {\bibinfo {title} {Local measures of dynamical quantum phase
  transitions},\ }\href {https://doi.org/10.1103/PhysRevB.104.075130}
  {\bibfield  {journal} {\bibinfo  {journal} {Phys. Rev. B}\ }\textbf {\bibinfo
  {volume} {104}},\ \bibinfo {pages} {075130} (\bibinfo {year}
  {2021})}\BibitemShut {NoStop}%
\bibitem [{\citenamefont {Bandyopadhyay}\ \emph {et~al.}(2021)\citenamefont
  {Bandyopadhyay}, \citenamefont {Polkovnikov},\ and\ \citenamefont
  {Dutta}}]{Bandyopadhyay2021}%
  \BibitemOpen
  \bibfield  {author} {\bibinfo {author} {\bibfnamefont {S.}~\bibnamefont
  {Bandyopadhyay}}, \bibinfo {author} {\bibfnamefont {A.}~\bibnamefont
  {Polkovnikov}},\ and\ \bibinfo {author} {\bibfnamefont {A.}~\bibnamefont
  {Dutta}},\ }\bibfield  {title} {\bibinfo {title} {Observing dynamical quantum
  phase transitions through quasilocal string operators},\ }\href
  {https://doi.org/10.1103/PhysRevLett.126.200602} {\bibfield  {journal}
  {\bibinfo  {journal} {Phys. Rev. Lett.}\ }\textbf {\bibinfo {volume} {126}},\
  \bibinfo {pages} {200602} (\bibinfo {year} {2021})}\BibitemShut {NoStop}%
\bibitem [{\citenamefont {Schmitt}\ and\ \citenamefont
  {Heyl}(2018)}]{schmittHeyl2018}%
  \BibitemOpen
  \bibfield  {author} {\bibinfo {author} {\bibfnamefont {M.}~\bibnamefont
  {Schmitt}}\ and\ \bibinfo {author} {\bibfnamefont {M.}~\bibnamefont {Heyl}},\
  }\bibfield  {title} {\bibinfo {title} {{Quantum dynamics in transverse-field
  Ising models from classical networks}},\ }\href
  {https://doi.org/10.21468/SciPostPhys.4.2.013} {\bibfield  {journal}
  {\bibinfo  {journal} {SciPost Phys.}\ }\textbf {\bibinfo {volume} {4}},\
  \bibinfo {pages} {013} (\bibinfo {year} {2018})}\BibitemShut {NoStop}%
\bibitem [{\citenamefont {Canovi}\ \emph
  {et~al.}(2014{\natexlab{b}})\citenamefont {Canovi}, \citenamefont
  {Ercolessi}, \citenamefont {Naldesi}, \citenamefont {Taddia},\ and\
  \citenamefont {Vodola}}]{CanoviPRB2014}%
  \BibitemOpen
  \bibfield  {author} {\bibinfo {author} {\bibfnamefont {E.}~\bibnamefont
  {Canovi}}, \bibinfo {author} {\bibfnamefont {E.}~\bibnamefont {Ercolessi}},
  \bibinfo {author} {\bibfnamefont {P.}~\bibnamefont {Naldesi}}, \bibinfo
  {author} {\bibfnamefont {L.}~\bibnamefont {Taddia}},\ and\ \bibinfo {author}
  {\bibfnamefont {D.}~\bibnamefont {Vodola}},\ }\bibfield  {title} {\bibinfo
  {title} {Dynamics of entanglement entropy and entanglement spectrum crossing
  a quantum phase transition},\ }\href
  {https://doi.org/10.1103/PhysRevB.89.104303} {\bibfield  {journal} {\bibinfo
  {journal} {Phys. Rev. B}\ }\textbf {\bibinfo {volume} {89}},\ \bibinfo
  {pages} {104303} (\bibinfo {year} {2014}{\natexlab{b}})}\BibitemShut
  {NoStop}%
\bibitem [{\citenamefont {Surace}\ \emph {et~al.}(2020)\citenamefont {Surace},
  \citenamefont {Tagliacozzo},\ and\ \citenamefont {Tonni}}]{Surace2020}%
  \BibitemOpen
  \bibfield  {author} {\bibinfo {author} {\bibfnamefont {J.}~\bibnamefont
  {Surace}}, \bibinfo {author} {\bibfnamefont {L.}~\bibnamefont
  {Tagliacozzo}},\ and\ \bibinfo {author} {\bibfnamefont {E.}~\bibnamefont
  {Tonni}},\ }\bibfield  {title} {\bibinfo {title} {Operator content of
  entanglement spectra in the transverse field {Ising chain} after global
  quenches},\ }\href {https://doi.org/10.1103/PhysRevB.101.241107} {\bibfield
  {journal} {\bibinfo  {journal} {Phys. Rev. B}\ }\textbf {\bibinfo {volume}
  {101}},\ \bibinfo {pages} {241107} (\bibinfo {year} {2020})}\BibitemShut
  {NoStop}%
\bibitem [{\citenamefont {P{\"o}yh{\"o}nen}\ and\ \citenamefont
  {Ojanen}(2021)}]{poyhonen2021entanglement}%
  \BibitemOpen
  \bibfield  {author} {\bibinfo {author} {\bibfnamefont {K.}~\bibnamefont
  {P{\"o}yh{\"o}nen}}\ and\ \bibinfo {author} {\bibfnamefont {T.}~\bibnamefont
  {Ojanen}},\ }\href@noop {} {\bibinfo {title} {Entanglement echo and dynamical
  entanglement transitions}} (\bibinfo {year} {2021}),\ \Eprint
  {https://arxiv.org/abs/2106.10043} {arXiv:2106.10043 [quant-ph]} \BibitemShut
  {NoStop}%
\bibitem [{\citenamefont {Halimeh}\ \emph {et~al.}(2020)\citenamefont
  {Halimeh}, \citenamefont {Van~Damme}, \citenamefont {Zauner-Stauber},\ and\
  \citenamefont {Vanderstraeten}}]{Halimeh2020}%
  \BibitemOpen
  \bibfield  {author} {\bibinfo {author} {\bibfnamefont {J.~C.}\ \bibnamefont
  {Halimeh}}, \bibinfo {author} {\bibfnamefont {M.}~\bibnamefont {Van~Damme}},
  \bibinfo {author} {\bibfnamefont {V.}~\bibnamefont {Zauner-Stauber}},\ and\
  \bibinfo {author} {\bibfnamefont {L.}~\bibnamefont {Vanderstraeten}},\
  }\bibfield  {title} {\bibinfo {title} {Quasiparticle origin of dynamical
  quantum phase transitions},\ }\href
  {https://doi.org/10.1103/PhysRevResearch.2.033111} {\bibfield  {journal}
  {\bibinfo  {journal} {Phys. Rev. Research}\ }\textbf {\bibinfo {volume}
  {2}},\ \bibinfo {pages} {033111} (\bibinfo {year} {2020})}\BibitemShut
  {NoStop}%
\bibitem [{\citenamefont {{Hogan}}\ and\ \citenamefont
  {{Chalker}}(2004)}]{hoganChalker}%
  \BibitemOpen
  \bibfield  {author} {\bibinfo {author} {\bibfnamefont {P.~M.}\ \bibnamefont
  {{Hogan}}}\ and\ \bibinfo {author} {\bibfnamefont {J.~T.}\ \bibnamefont
  {{Chalker}}},\ }\bibfield  {title} {\bibinfo {title} {{Path integrals,
  diffusion on SU(2) and the fully frustrated antiferromagnetic spin
  cluster}},\ }\href {https://doi.org/10.1088/0305-4470/37/49/002} {\bibfield
  {journal} {\bibinfo  {journal} {J. Phys. A: Math. Gen.}\ }\textbf {\bibinfo
  {volume} {37}},\ \bibinfo {pages} {11751} (\bibinfo {year}
  {2004})}\BibitemShut {NoStop}%
\bibitem [{\citenamefont {Ringel}\ and\ \citenamefont
  {Gritsev}(2013)}]{ringelGritsev}%
  \BibitemOpen
  \bibfield  {author} {\bibinfo {author} {\bibfnamefont {M.}~\bibnamefont
  {Ringel}}\ and\ \bibinfo {author} {\bibfnamefont {V.}~\bibnamefont
  {Gritsev}},\ }\bibfield  {title} {\bibinfo {title} {Dynamical symmetry
  approach to path integrals of quantum spin systems},\ }\href
  {https://doi.org/10.1103/PhysRevA.88.062105} {\bibfield  {journal} {\bibinfo
  {journal} {Phys. Rev. A}\ }\textbf {\bibinfo {volume} {88}},\ \bibinfo
  {pages} {062105} (\bibinfo {year} {2013})}\BibitemShut {NoStop}%
\bibitem [{\citenamefont {{De Nicola}}\ \emph {et~al.}(2019)\citenamefont {{De
  Nicola}}, \citenamefont {Doyon},\ and\ \citenamefont
  {Bhaseen}}]{stochasticApproach}%
  \BibitemOpen
  \bibfield  {author} {\bibinfo {author} {\bibfnamefont {S.}~\bibnamefont {{De
  Nicola}}}, \bibinfo {author} {\bibfnamefont {B.}~\bibnamefont {Doyon}},\ and\
  \bibinfo {author} {\bibfnamefont {M.~J.}\ \bibnamefont {Bhaseen}},\
  }\bibfield  {title} {\bibinfo {title} {{Stochastic approach to
  non-equilibrium quantum spin systems}},\ }\href
  {https://doi.org/10.1088/1751-8121/aaf9be} {\bibfield  {journal} {\bibinfo
  {journal} {J. Phys. A: Math. Theor.}\ }\textbf {\bibinfo {volume} {52}},\
  \bibinfo {pages} {05LT02} (\bibinfo {year} {2019})}\BibitemShut {NoStop}%
\bibitem [{\citenamefont {{De Nicola}}\ \emph {et~al.}(2020)\citenamefont {{De
  Nicola}}, \citenamefont {Doyon},\ and\ \citenamefont
  {Bhaseen}}]{nonEquilibrium}%
  \BibitemOpen
  \bibfield  {author} {\bibinfo {author} {\bibfnamefont {S.}~\bibnamefont {{De
  Nicola}}}, \bibinfo {author} {\bibfnamefont {B.}~\bibnamefont {Doyon}},\ and\
  \bibinfo {author} {\bibfnamefont {M.~J.}\ \bibnamefont {Bhaseen}},\
  }\bibfield  {title} {\bibinfo {title} {Non-equilibrium quantum spin dynamics
  from classical stochastic processes},\ }\href
  {https://doi.org/10.1088/1742-5468/ab6093} {\bibfield  {journal} {\bibinfo
  {journal} {J. Stat. Mech.: Theory Exp.}\ }\textbf {\bibinfo {volume}
  {2020}}\bibinfo  {number} { (1)},\ \bibinfo {pages} {013106}}\BibitemShut
  {NoStop}%
\bibitem [{\citenamefont {{De Nicola}}(2021)}]{sdn2021}%
  \BibitemOpen
\bibfield  {number} {  }\bibfield  {author} {\bibinfo {author} {\bibfnamefont
  {S.}~\bibnamefont {{De Nicola}}},\ }\href@noop {} {\bibinfo {title}
  {Importance sampling scheme for the stochastic simulation of quantum spin
  dynamics}} (\bibinfo {year} {2021}),\ \Eprint
  {https://arxiv.org/abs/2103.16468} {arXiv:2103.16468 [quant-ph]} \BibitemShut
  {NoStop}%
\bibitem [{\citenamefont {Schuch}\ \emph {et~al.}(2007)\citenamefont {Schuch},
  \citenamefont {Wolf}, \citenamefont {Verstraete},\ and\ \citenamefont
  {Cirac}}]{Schuch2007}%
  \BibitemOpen
  \bibfield  {author} {\bibinfo {author} {\bibfnamefont {N.}~\bibnamefont
  {Schuch}}, \bibinfo {author} {\bibfnamefont {M.~M.}\ \bibnamefont {Wolf}},
  \bibinfo {author} {\bibfnamefont {F.}~\bibnamefont {Verstraete}},\ and\
  \bibinfo {author} {\bibfnamefont {J.~I.}\ \bibnamefont {Cirac}},\ }\bibfield
  {title} {\bibinfo {title} {Computational complexity of projected entangled
  pair states},\ }\href {https://doi.org/10.1103/PhysRevLett.98.140506}
  {\bibfield  {journal} {\bibinfo  {journal} {Phys. Rev. Lett.}\ }\textbf
  {\bibinfo {volume} {98}},\ \bibinfo {pages} {140506} (\bibinfo {year}
  {2007})}\BibitemShut {NoStop}%
\bibitem [{\citenamefont {Czarnik}\ \emph {et~al.}(2019)\citenamefont
  {Czarnik}, \citenamefont {Dziarmaga},\ and\ \citenamefont
  {Corboz}}]{Czarnik2019}%
  \BibitemOpen
  \bibfield  {author} {\bibinfo {author} {\bibfnamefont {P.}~\bibnamefont
  {Czarnik}}, \bibinfo {author} {\bibfnamefont {J.}~\bibnamefont {Dziarmaga}},\
  and\ \bibinfo {author} {\bibfnamefont {P.}~\bibnamefont {Corboz}},\
  }\bibfield  {title} {\bibinfo {title} {Time evolution of an infinite
  projected entangled pair state: An efficient algorithm},\ }\href
  {https://doi.org/10.1103/PhysRevB.99.035115} {\bibfield  {journal} {\bibinfo
  {journal} {Phys. Rev. B}\ }\textbf {\bibinfo {volume} {99}},\ \bibinfo
  {pages} {035115} (\bibinfo {year} {2019})}\BibitemShut {NoStop}%
\bibitem [{\citenamefont {Srivastav}\ \emph {et~al.}(2019)\citenamefont
  {Srivastav}, \citenamefont {Bhattacharya},\ and\ \citenamefont
  {Dutta}}]{srivastav2019}%
  \BibitemOpen
  \bibfield  {author} {\bibinfo {author} {\bibfnamefont {V.}~\bibnamefont
  {Srivastav}}, \bibinfo {author} {\bibfnamefont {U.}~\bibnamefont
  {Bhattacharya}},\ and\ \bibinfo {author} {\bibfnamefont {A.}~\bibnamefont
  {Dutta}},\ }\bibfield  {title} {\bibinfo {title} {Dynamical quantum phase
  transitions in extended toric-code models},\ }\href
  {https://doi.org/10.1103/PhysRevB.100.144203} {\bibfield  {journal} {\bibinfo
   {journal} {Phys. Rev. B}\ }\textbf {\bibinfo {volume} {100}},\ \bibinfo
  {pages} {144203} (\bibinfo {year} {2019})}\BibitemShut {NoStop}%
\bibitem [{\citenamefont {Bhattacharya}\ and\ \citenamefont
  {Dutta}(2017)}]{bhattacharya2017}%
  \BibitemOpen
  \bibfield  {author} {\bibinfo {author} {\bibfnamefont {U.}~\bibnamefont
  {Bhattacharya}}\ and\ \bibinfo {author} {\bibfnamefont {A.}~\bibnamefont
  {Dutta}},\ }\bibfield  {title} {\bibinfo {title} {Emergent topology and
  dynamical quantum phase transitions in two-dimensional closed quantum
  systems},\ }\href {https://doi.org/10.1103/PhysRevB.96.014302} {\bibfield
  {journal} {\bibinfo  {journal} {Phys. Rev. B}\ }\textbf {\bibinfo {volume}
  {96}},\ \bibinfo {pages} {014302} (\bibinfo {year} {2017})}\BibitemShut
  {NoStop}%
\bibitem [{\citenamefont {Onsager}(1944)}]{Onsager44}%
  \BibitemOpen
  \bibfield  {author} {\bibinfo {author} {\bibfnamefont {L.}~\bibnamefont
  {Onsager}},\ }\bibfield  {title} {\bibinfo {title} {Crystal statistics. {I}.
  {A} two-dimensional model with an order-disorder transition},\ }\href
  {https://doi.org/10.1103/PhysRev.65.117} {\bibfield  {journal} {\bibinfo
  {journal} {Phys. Rev.}\ }\textbf {\bibinfo {volume} {65}},\ \bibinfo {pages}
  {117} (\bibinfo {year} {1944})}\BibitemShut {NoStop}%
\bibitem [{\citenamefont {Heyl}(2015)}]{heyl2015}%
  \BibitemOpen
  \bibfield  {author} {\bibinfo {author} {\bibfnamefont {M.}~\bibnamefont
  {Heyl}},\ }\bibfield  {title} {\bibinfo {title} {Scaling and universality at
  dynamical quantum phase transitions},\ }\href
  {https://doi.org/10.1103/PhysRevLett.115.140602} {\bibfield  {journal}
  {\bibinfo  {journal} {Phys. Rev. Lett.}\ }\textbf {\bibinfo {volume} {115}},\
  \bibinfo {pages} {140602} (\bibinfo {year} {2015})}\BibitemShut {NoStop}%
\bibitem [{\citenamefont {James}\ and\ \citenamefont
  {Konik}(2015)}]{James2015}%
  \BibitemOpen
  \bibfield  {author} {\bibinfo {author} {\bibfnamefont {A.~J.~A.}\
  \bibnamefont {James}}\ and\ \bibinfo {author} {\bibfnamefont {R.~M.}\
  \bibnamefont {Konik}},\ }\bibfield  {title} {\bibinfo {title} {Quantum
  quenches in two spatial dimensions using chain array matrix product states},\
  }\href {https://doi.org/10.1103/PhysRevB.92.161111} {\bibfield  {journal}
  {\bibinfo  {journal} {Phys. Rev. B}\ }\textbf {\bibinfo {volume} {92}},\
  \bibinfo {pages} {161111} (\bibinfo {year} {2015})}\BibitemShut {NoStop}%
\bibitem [{\citenamefont {Hashizume}\ \emph {et~al.}(2018)\citenamefont
  {Hashizume}, \citenamefont {McCulloch},\ and\ \citenamefont
  {Halimeh}}]{Hashizume2020}%
  \BibitemOpen
  \bibfield  {author} {\bibinfo {author} {\bibfnamefont {T.}~\bibnamefont
  {Hashizume}}, \bibinfo {author} {\bibfnamefont {I.~P.}\ \bibnamefont
  {McCulloch}},\ and\ \bibinfo {author} {\bibfnamefont {J.~C.}\ \bibnamefont
  {Halimeh}},\ }\href@noop {} {\bibinfo {title} {Dynamical phase transitions in
  the two-dimensional transverse-field ising model}} (\bibinfo {year} {2018}),\
  \Eprint {https://arxiv.org/abs/1811.09275} {arXiv:1811.09275
  [cond-mat.str-el]} \BibitemShut {NoStop}%
\bibitem [{\citenamefont {Vidal}(2007)}]{Vidal2006}%
  \BibitemOpen
  \bibfield  {author} {\bibinfo {author} {\bibfnamefont {G.}~\bibnamefont
  {Vidal}},\ }\bibfield  {title} {\bibinfo {title} {Classical simulation of
  infinite-size quantum lattice systems in one spatial dimension},\ }\href
  {https://doi.org/10.1103/PhysRevLett.98.070201} {\bibfield  {journal}
  {\bibinfo  {journal} {Phys. Rev. Lett.}\ }\textbf {\bibinfo {volume} {98}},\
  \bibinfo {pages} {070201} (\bibinfo {year} {2007})}\BibitemShut {NoStop}%
\bibitem [{\citenamefont {Or\'us}\ and\ \citenamefont
  {Vidal}(2008)}]{Orus2008}%
  \BibitemOpen
  \bibfield  {author} {\bibinfo {author} {\bibfnamefont {R.}~\bibnamefont
  {Or\'us}}\ and\ \bibinfo {author} {\bibfnamefont {G.}~\bibnamefont {Vidal}},\
  }\bibfield  {title} {\bibinfo {title} {Infinite time-evolving block
  decimation algorithm beyond unitary evolution},\ }\href
  {https://doi.org/10.1103/PhysRevB.78.155117} {\bibfield  {journal} {\bibinfo
  {journal} {Phys. Rev. B}\ }\textbf {\bibinfo {volume} {78}},\ \bibinfo
  {pages} {155117} (\bibinfo {year} {2008})}\BibitemShut {NoStop}%
\bibitem [{\citenamefont {Piroli}\ \emph {et~al.}(2018)\citenamefont {Piroli},
  \citenamefont {Pozsgay},\ and\ \citenamefont {Vernier}}]{piroli2018}%
  \BibitemOpen
  \bibfield  {author} {\bibinfo {author} {\bibfnamefont {L.}~\bibnamefont
  {Piroli}}, \bibinfo {author} {\bibfnamefont {B.}~\bibnamefont {Pozsgay}},\
  and\ \bibinfo {author} {\bibfnamefont {E.}~\bibnamefont {Vernier}},\
  }\bibfield  {title} {\bibinfo {title} {{Non-analytic behavior of the
  Loschmidt echo in XXZ spin chains: Exact results}},\ }\href
  {https://doi.org/https://doi.org/10.1016/j.nuclphysb.2018.06.015} {\bibfield
  {journal} {\bibinfo  {journal} {Nucl. Phys. B.}\ }\textbf {\bibinfo {volume}
  {933}},\ \bibinfo {pages} {454 } (\bibinfo {year} {2018})}\BibitemShut
  {NoStop}%
\bibitem [{\citenamefont {Tamascelli}\ \emph {et~al.}(2015)\citenamefont
  {Tamascelli}, \citenamefont {Rosenbach},\ and\ \citenamefont
  {Plenio}}]{rSVD2015}%
  \BibitemOpen
  \bibfield  {author} {\bibinfo {author} {\bibfnamefont {D.}~\bibnamefont
  {Tamascelli}}, \bibinfo {author} {\bibfnamefont {R.}~\bibnamefont
  {Rosenbach}},\ and\ \bibinfo {author} {\bibfnamefont {M.~B.}\ \bibnamefont
  {Plenio}},\ }\bibfield  {title} {\bibinfo {title} {Improved scaling of
  time-evolving block-decimation algorithm through reduced-rank randomized
  singular value decomposition},\ }\href
  {https://doi.org/10.1103/PhysRevE.91.063306} {\bibfield  {journal} {\bibinfo
  {journal} {Phys. Rev. E}\ }\textbf {\bibinfo {volume} {91}},\ \bibinfo
  {pages} {063306} (\bibinfo {year} {2015})}\BibitemShut {NoStop}%
\end{thebibliography}
\end{document}